\documentclass{aastex631}

\shorttitle{Critical Mach Numbers for MHD Shocks}
\shortauthors{Laming}
\graphicspath{{./}{figures/}}

\begin{document}

\title{Critical Mach Numbers for Magnetohydrodynamic Shocks with Accelerated Particles and Waves}

\correspondingauthor{J. Martin Laming}
\email{laming@nrl.navy.mil}

\author[0000-0002-3362-7040]{J. Martin Laming}
\affiliation{Space Science Division, Code 7684, Naval Research Laboratory, 
Washington DC 20375, USA}



\begin{abstract}

The first critical fast Mach number is defined for a magnetohydrodynamic shock
as the Mach number where the shock transitions from subcritical, laminar, behavior 
to supercritical behavior, characterized by incident ion reflection from the shock front.
The ensuing upstream waves and turbulence are convected downstream leading to a turbulent shock
structure. Formally this is the Mach number where plasma resistivity
can no longer provide sufficient dissipation to establish a stable shock, and is characterized by the 
downstream flow speed becoming subsonic. We revisit these
calculations, including in the MHD jump conditions terms modeling the plasma energy loss to 
accelerated particles and the presence of waves associated with these particles. 
The accelerated particle contributions make an insignificant change, but the associated waves have a
more important effect. Upstream waves can be strongly amplified in intensity on passing through
the shock, and represent another means of shock dissipation. The presence of such waves therefore
increases the first critical fast Mach number, especially at quasi-parallel shock where wave excitation
is strongest. These effects may have significance for the solar regions where shock waves accelerate
particles and cause Type II and Type III radio bursts, and also contribute to the event-to-event variability of 
SEP acceleration.

\end{abstract}

\section{Introduction} \label{sec:intro}
Solar Energetic Particles (SEPs) accelerated at shock waves driven by Coronal Mass Ejections (CMEs) comprise the most serious space weather hazard for
instrumentation (and humans) in orbit outside the protection of the Earth's magnetic field. In the most extreme and damaging of these events, SEP
acceleration is inferred to begin close to the Sun, almost as soon as the solar wind disturbance initiated by the CME steepens into a shock. A common
assumption is to connect the onset of SEP acceleration with the magnetohydrodynamic (MHD) 
shock strengthening beyond the so-called ``first critical Mach number'' \citep[e.g.][]{mann95,lepri12}, following the suggestion of \citet{edmiston84}. 
This point marks where dissipation beyond that provided by the plasma resistivity
is required to establish the shock structure \citep{edmiston84}, and can be identified with conditions where the
postshock flow speed becomes subsonic \citep{coroniti70}. On transitioning from subcritical to supercritical,
the shock ceases to be laminar and a shock precursor of reflected ions develops \citep{treumann09}. In this way waves and turbulence can develop upstream,
which are necessary conditions for energetic particle acceleration. In fact this argument can be turned around to say
that supercritical collisionless shock waves {\em must} reflect ions \citep{treumann08}, and therefore that energetic particle acceleration becomes
inevitable in order to achieve sufficient dissipation in the shock. 

The first critical fast Mach number is also crucial to other phenomena, such as Type II and Type III radio bursts, where
shock generated turbulence is required to accelerate electron beams 
and might also have consequences for the shock morphology observed by imaging instruments. 
According to \citet{ramesh22} and references therein, Type II radio bursts usually propagate at Alfv\'en Mach numbers $M_A =v/v_A < 2$, which implies
association with a shock of similar $M_A$. 
\citet{mann95} had previously argued that such observations suggest either supercritical quasi-parallel or subcritical quasi-perpendicular shocks, 
with a possible threshold at shock compression $r\simeq 1.35$ favoring supercritical quasi-parallel shocks. Similarly,
\citet{maguire20} favor a threshold of $M_A = 1.4 - 2.4$ for Type II burst onset, implying a supercritical quasi-parallel or 
quasi-perpendicular shock, and a burst cessation 
when the shock reaches $\sim 2.4$ R$_{\sun}$ and becomes closer to being strictly parallel diminishing efficient electron
acceleration as magnetic field lines become more radial.

While Type II radio bursts seem to require supercritical shocks,  the observations of 
\citet{zhu18} imply that SEPs can be accelerated while the shock is subcritical. Studies of the 2012 January 17 event reveal that SEPs accelerated
by the shock are {\em released} when shock hits the first critical Mach number. \citet{zhu18} determine $M_A\simeq 1.5$ at release, implying SEP
acceleration occurring for $1 < M_A < 1.5$. This is potentially an important insight with implications for predicting the onset and severity of SEP
events, and matches with recent suggestions that SEP events are favored when the CME driven shock encounters a population of already suprathermal ions 
that can act as seed particles for the acceleration process \citep[e.g.][and references therein]{laming13}. In such a case, the shock itself is not required
to generate its own turbulence (by becoming supercritical), because the pre-existing seed particle distribution does that upon interacting with the shock.
Type II bursts then also require energetic particle escape from the shock for onset, similarly to Type III bursts, although in the Type II case, the 
released particles do not propagate far ahead. The recent studies of \citet{gopalswamy16}, \citet{pesce22} and \citet{klein22}
can also be interpreted in this way. Gamma ray emission 
from a behind-the-limb solar flare is detected by the Fermi Large Area Telescope (LAT) in coincidence with a Type II radio burst. 
Direct gamma rays from the flare site are of course occulted, so the Fermi-LAT emission must be due to particles accelerated at a shock associated
with the flare, escaping and impinging on the chromosphere on the near side of the Sun. These particles are probably escaping from the rear of the shock, towards to downstream medium, while those that escape towards the upstream excite Langmuir waves and cause the Type II radio burst.

The calculations of \citet{edmiston84} are generally referred to in considerations of shock criticality, but in the case that the shock has already begun
accelerating SEPs out of a pre-existing seed particle population, the shock jump conditions will be modified by the presence of SEPs and their associated waves and turbulence. In the following sections we evaluate this effect. Section 2 discusses the revised shock jump conditions, while section 3 solves for the 
density compression and hence derives the first critical fast Mach number as a function of plasma $\beta$ (the ratio of gas pressure to magnetic pressure) 
and the shock obliquity. Section 4 offers some discussion and conclusions, most important of which will be that the first critical fast Mach number {\em increases} with
the presence of pre-existing accelerated particles and waves. In the event that the shock transition to supercriticality is the primary step towards release
of SEPs, this effect will mean that a shock strongly loaded with accelerated particles will being SEP release later rather than sooner, and at presumably
higher SEP energy too, making the first critical fast Mach number and its variability with SEP and wave pressure an important parameter controlling the
variation in the SEP effectiveness of CME driven shocks. Much of the literature concerning particle acceleration at shocks focuses on astrophysical
rather than heliophysical settings, e.g. shocks driven by supernova remnants, with the accelerated particles referred to as ``cosmic rays'' rather than
SEPs. Throughout this paper we use both terms interchangeably, depending mainly on the literature being cited. Some of the more technical details concerning wave reflection and transmission coefficients at oblique MHD shocks, and the presence
(or absence) of other critical Mach numbers in the flow are given in Appendices A and B respectively.

\section{Shock Jump Conditions} \label{sec:jump}
Historically, the effect of cosmic rays on shock discontinuities has been treated in the two-fluid hydrodynamic approximation 
\citep[e.g.][]{drury81,drury83} with a more comprehensive analysis of the various important Mach numbers involved given by \citet{becker01}.
These treatments have separate pressure
and energy fluxes for the thermal gas and the cosmic rays, treated as fluids with adiabatic indices $\gamma =5/3$ and 4/3 respectively. Magnetic fields have not been included, nor in general have waves or turbulence. In a review, \citet{drury83} summarizes prior work on including waves in hydrodynamical
models, and \citet{ko92} gives further discussion, with the inclusion of forward and backward going waves and second order Fermi acceleration.
In contrast to the models above with pre-existing cosmic rays, 
\citet{vink14} investigate a two-fluid model of a cosmic ray accelerating shock, on the basis of which they derive a critical sonic Mach number $M_s = \sqrt{5}$ as a threshold for particle acceleration, or more precisely, a threshold for the development of a shock precursor associated with particle
acceleration. Exceptions to this occur when there are pre-existing energetic particles.  

\begin{figure}[t]
\centerline{\includegraphics[width=3.in]{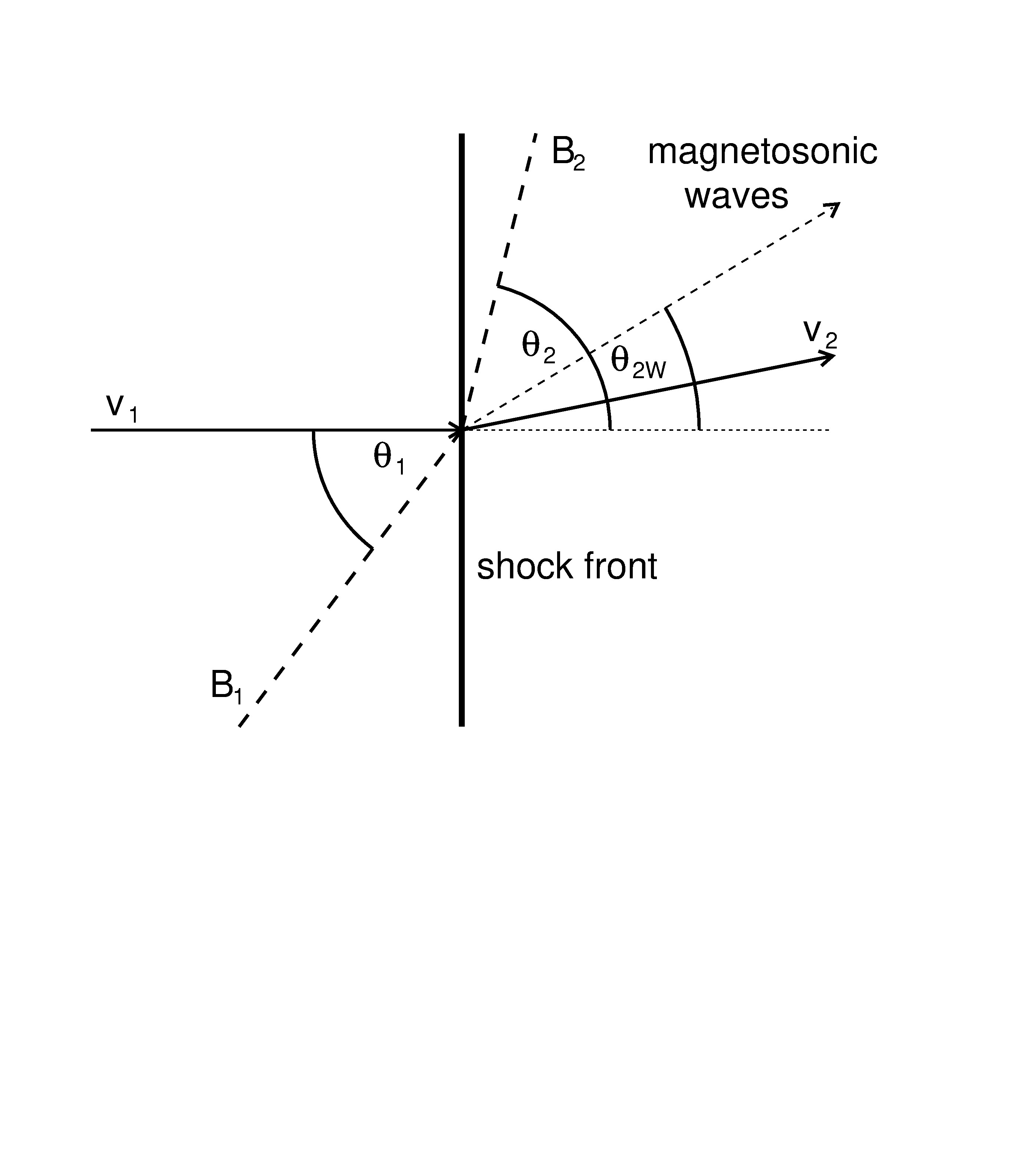}\includegraphics[width=3.in]{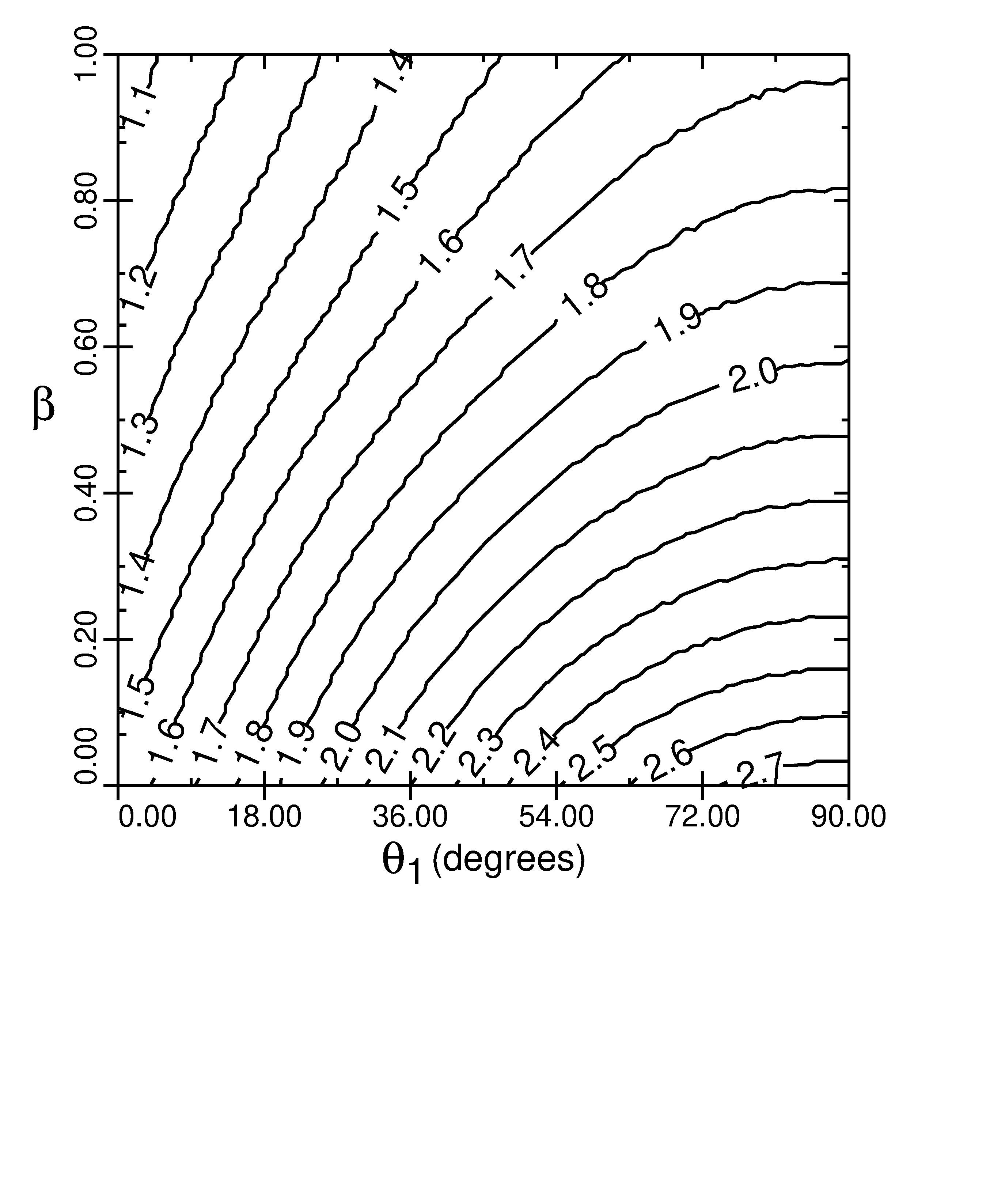}}
\vspace{-0.9truein}
\caption{Left: Schematic diagram of wave propagation at the oblique shock, in the shock rest frame. Upstream (left), plasma moves towards the shock front with velocity ${\bf u}_1$, carrying with it magnetic field ${\bf B}_1$ at angle $\theta _1$ to ${\bf v}_1$ (or the shock normal). Postshock, the magnetic field is ${\bf B}_2$ at angle $\theta _2$ to the shock normal and the flow velocity is ${\bf v}_2$. In the preshock medium, waves are assumed to be parallel propagating along ${\bf B}_1$. Postshock, the magnetosonic waves refract and travel at an angle $\theta _{2W}$ to the shock normal. Alfv\'en waves
refract at a slightly different angle, but this plays no role in their reflection and transmission coefficients. 
Right: Contour plot of the first critical fast Mach number in the plasma $\beta$ - $\theta _1$ plane for $E_{CR}/\rho _1v_{1||}^2 = 0$ 
\citep[see Fig 4c in][]{edmiston84}. \label{shock}}
\end{figure}

We consider jump conditions for the thermal gas at an MHD shock, including contributions from magnetic field and MHD waves. Cosmic rays,
or SEPs, obey separate jump conditions \citep{drury81,drury83,becker01}. They only appear as sinks of energy and momentum in the thermal gas jump
conditions. When the cosmic ray number density $n_{CR}<<n$, the thermal
gas number density, the cosmic ray contribution to the mass continuity can be neglected. The cosmic ray pressure is continuous across the
shock, so they make no contribution to the momentum jump. And working in the rest frame of the shock, in the 
absence of cosmic ray escape, we take the cosmic ray energy flux to be zero, so those terms also drop out of the energy equation. We emphasize that 
our goal is not to model all details of a cosmic ray modified shock, but more simply to estimate the effects of accelerated particles and waves on the 
shock criticality.

A schematic diagram of the MHD shock is given in the left panel of Fig. \ref{shock}. The upstream magnetic field ${\bf B}_1$ makes an angle
$\theta _1$ to the shock normal. A frame of reference is chosen so that plasma inflow to the shock with velocity ${\bf v}_1$  is along this shock normal.
Postshock the magnetic field ${\bf B}_2$ makes an angle $\theta _2$ to the shock normal, and the postshock flow speed ${\bf v}_2$ has 
normal and transverse components. The waves associated with the SEPs are generated in the upstream medium and assumed to be parallel
propagating with respect to ${\bf B}_1$. These are amplified and refracted upon passage through the shock. The amplification is treated by
calculating the reflection and transmission coefficients for forward and backward propagating waves at an oblique shock, generalizing the earlier
case considered by \citet{laming15} of high Mach number to arbitrary Mach number. In the case of Alfv\'en polarization, the wave refraction plays no role
in the transmission and reflection coefficients, because the wave magnetic field is perpendicular to the plane defined by the wavevectors in 
upstream and downstream. For the magnetosonic polarization, this angle of refraction, denoted here as $\theta _{2w}$ with respect to the shock
normal, does have an effect on the transmission and reflection coefficients. 
It is interesting to note that the Alfv\'en and magnetosonic polarizations in fact refract at {\em different} angles, so
an initially circularly polarized upstream wave becomes decomposed into its plane polarized Alfv\'en and magnetosonic components downstream. 
This difference tends to zero as the Alfv\'en Mach number becomes infinite. 
The various relations between upstream and downstream shock parameters, and the wave reflection and transmission
coefficients of relevance to this paper are collected in Appendix A.

With subscripts 1 and 2 denoting the upstream and downstream regions respectively, the mass jump condition is written
\begin{equation}
\rho _1v_{1||}=\rho _2v_{2||}
\end{equation}
for densities $\rho _{1,2}$, velocities $v_{1,2}$ and contains nothing remarkable. We write the thermal gas momentum jump condition as
\begin{equation}
\rho _1v_{1||}^2+P_1+P_{1w}+\frac{B_{1\perp}^2}{8\pi}+F_1v_{1||}=
\rho _2v_{2||}^2+P_2+P_{2w}+\frac{B_{2\perp}^2}{8\pi}+F_2v_{2||}
\end{equation}
for pressures {\bf $P_1$ and $P_2$} and magnetic fields
{\bf $B_1$ and $B_2$} in upstream and downstream respectively. In equation 2,
$F_1=F_2=-\left(2/3\right)n_{CR}p_{CR}\left(v_{1||}-v_{2||}\right)/\left<v_{CR}\right>$ are the momenta imparted 
to the population of accelerated particles in terms of their number density, $n_{CR}$ and momenta $p_{CR}$,
by the diffusive shock acceleration process in the upstream and downstream regions \citep[e.g.][section 2.3.2, equation 2.47]{drury83}. We take
$1/\left<v_{CR}\right> = \sum _i1/v_i/n_{CR}$ where the $v_i$ are the individual cosmic ray velocities and $n_{CR}$ is their number density, which is 
clearly dominated by the lowest energy cosmic rays which can be injected into the acceleration process.
Working in the shock rest frame, at the shock, the upstream medium
has lost momentum to the accelerated particles relative to the far upstream. Downstream, some of this momentum is gained back. 

In our case $P_{1w}$ and $P_{2w}$ represent
the combined Alfv\'en and magnetosonic wave pressures in upstream and downstream, coming from forward and backward propagating waves. These 
are connected by the amplitude transmission $T$ and reflection $R$ coefficients at the shock as follows
\begin{equation}
P_{2w}=P_{2wf}+P_{2wb}=P_{1wf}T_f^2+P_{1wb}R_b^2+P_{1wb}T_b^2+P_{1wf}R_f^2.
\end{equation}
The subscripts $f$ and $b$ refer to 
forward and backward propagating waves with magnetic perturbations $\delta B_{1b}$, $\delta B_{1f}$ and similarly for region 2. The transmission
and reflection coefficients are defined by $T_f=\delta B_{2f}/\delta B_{1f}$, $R_f=\delta B_{2b}/\delta B_{1f}$, 
$T_b=\delta B_{2b}/\delta B_{1b}$, and $R_b=\delta B_{2f}/\delta B_{1b}$.
Full expressions for $T_{f,b}$ and $R_{f,b}$ derived from \citet{laming15} are
given in the Appendix. As $M_{A1}=v_{1||}/v_{A1}\rightarrow\infty$, $T_f=T_b=\left(r+\sqrt{r}\right)/2$, 
$R_f=R_b=\left(r-\sqrt{r}\right)/2$ for Alfv\'en polarization and 
$\left(r\pm\sqrt{r}\right)\sqrt{\sin ^2\theta_1+r^2\cos ^2\theta _1}/2$
for the magnetosonic polarization, where $r=\rho _2/\rho _1=v_{1||}/v_{2||}$ is the shock compression. 

Equation 2 implies a cosmic ray momentum equation $P_{1CR}-F_1v_{1CR}=P_{2CR}-F_2v_{2CR}$.  If $v_{1CR}=v_{1||}$
and $v_{2CR}=v_{2||}$, then equation 2 takes a form with $F_1v_{1||}\rightarrow P_{1CR}$ and $F_2v_{2||}\rightarrow P_{2CR}$.
However in the case we consider, where cosmic rays are being accelerated but are not yet escaping from the shock, $v_{1CR} < 0$ and $v_2{CR} > 2$, as
the trapped cosmic ray populations grows, increasing in distance both
ahead of and behind the shock. The magnitudes of both are $\left|v_{1CR}\right|<< v_{1||}$ and $v_{2CR}<<v_{2||}$. If these values are allowed
to tend to zero, i.e. equal numbers of cosmic rays propagating back and forth across the shock, then $P_{1CR}=P_{2CR}$.

The energy jump condition is 
\begin{eqnarray}
&\frac{1}{2}&\rho _1v_{1||}^3+\frac{\gamma}{\gamma -1}P_1v_{1||}+2P_{1w}v_{1||}+\frac{B_{1\perp}^2}{4\pi}v_{1||}+E_1v_{1||}=\cr
&\frac{1}{2}&\rho _2v_{2||}^3+\frac{1}{2}\rho _2v_{2\perp}^2v_{2||}+\frac{\gamma}{\gamma -1}P_2v_{2||}
+2P_{2w}v_{2||}+\frac{B_{2\perp}^2}{4\pi}v_{2||}+E_2v_{2||}
\end{eqnarray}
where $E_1 = E_2 = -\left(2/3\right)n_{CR}p_{CR}\left(v_{1||}-v_{2||}\right)$ are the energies imparted to the SEPs by the acceleration process. Again,
working in the rest frame of the shock, energy is absorbed by the accelerated particles from the upstream medium, and some is deposited back downstream.
In the rest frame of the upstream medium, energy and momentum are absorbed from the motion of the shocked gas, with some given back to the 
upstream. The wave energy flux terms in equation 4 assume equal intensities of forward and backward propagating waves in both the upstream 
and downstream media. This is the simplest case, and we discuss the relaxation of this approximation below in equations 10 and 11. Anticipating 
our final interpretation below, we ignore cosmic ray
escape from the acceleration process, so that the cosmic ray energy flux in the shock rest frame is close to zero. There are approximately
equal numbers of accelerated particles and equal energy fluxes
propagating in each direction across the shock in the diffusive shock acceleration process.

Prior treatments of purely hydrodynamic shocks \citep[e.g.][]{drury81,drury83,becker01} do not consider the terms in $F_1$, $F_2$, $E_1$, or $E_2$. In 
fact they make very little difference to our results below, being only discernable for the highest cosmic ray pressure case in Figure 2. 
We include them because they do capture the shock compression due to energy loss to cosmic
rays (see equations 8 and 11 below), which works in the opposite sense to the more dominant effect of the waves, and therefore cannot be neglected in all cases. These prior treatments
also assume cosmic rays drifting with the gas flow speed upstream and downstream, and not trapped at the shock in the acceleration process as we do.
These approximations are considered in more detail in Appendix B.

The wave intensity is given by the model of \citet{bell01}, generalized to oblique shocks. When $\delta B << B$, 
\begin{equation}
P_{1w}= P_{CR}\cos\theta _1/2M_A.
\end{equation} 
For larger amplitude waves, where $v_A$  is determined by $\delta B$, \citet{bell01} give expressions in the cases that either isotropization or advection
of waves dominates. These are appropriate at higher Mach numbers, $M_A > 4\rho _1v_{1||}^2/P_{CR}\cos\theta _1$ than are relevant in our study here.


We rearrange equation 2 in favor of $P_2$, and substitute into equation 4 to derive an equation for the shock compression, $r$. Expressions for
$B_{2\perp}$ and $v_{2\perp}$ are given in Appendix A, equations A1 and A2. We also assume that $P_{1w}$ is evenly split between Alfv\'en and
magnetosonic polarizations, and that both $P_{1w}$ and $P_{2w}$ have equal intensities of forward and backward propagating waves. We write 
$T^2=\sum _{pol}\left(T_f^2+T_b^2\right)/4$ and $R^2=\sum _{pol}\left(R_f^2+Rb^2\right)/4$, treating all waves as incoherent and neglecting
any interference, to find
\begin{eqnarray}
\bigg\{\frac{a}{r}-\frac{1}{M_{s1}^2}&+&\frac{P_{1w}}{\rho _1v_{1||}^2}\frac{\gamma -2r\left(\gamma -1\right)
-\left(T^2+R^2\right)\left(2-\gamma\right)}{r-1}
+\frac{F_1/r-F_2/r^2}{\rho _1v_{1||}}\frac{\gamma}{r -1}-\frac{E_1-E_2/r}{\rho_1v_{1||}^2}\frac{\gamma -1}{r-1}\bigg\}\left(\frac{1}{r}
-\frac{\cos ^2\theta _1}{M_{A1}^2}\right)^2 \cr
&-& \frac{\sin ^2\theta_1}{M_{A1}^2r}\left\{\frac{1}{r}\left(a-\frac{1-r}{2}\right)-\frac{a\cos ^2\theta_1}{M_{A1}^2}\right\}=0,
\end{eqnarray}
where $a=\left(\gamma +1\right)/2-r\left(\gamma -1\right)/2$.
In the absence of the cosmic ray terms, equation 5 reduces to the standard result 
\citep[e.g.][equation 8.75, with a typographical error corrected]{melrose86} or \citet[][equation 5.48]{priest14}.
For a parallel shock ($\theta _1=0^{\circ}$) equation 5 simplifies to 
\begin{equation}
\frac{\gamma +1}{2r}-\frac{\gamma -1}{2}-\frac{1}{M_{s1}^2}+\frac{P_{1w}}{\rho _1v_{1||}^2}\frac{\gamma -2r\left(\gamma -1\right)
-\left(T^2+R^2\right)\left(2-\gamma\right)}{r-1}
-\frac{2}{3}\frac{n_{CR}p_{CR}}{\rho _1v_{1||}}\left(r-1\right)\left\{\frac{v_{1||}}{v_{CR}}\frac{\gamma}{r^3}
-\frac{\gamma -1}{r^2}\right\}=0
\end{equation}
which, when $P_{1w}$ is neglected and $v_{CR} >> v_{1||}$, is a quadratic equation in $r$ with solution
\begin{eqnarray}
r&=&\frac{\frac{\gamma +1}{2}+\frac{2}{3}\frac{P_{CR}}{\rho _1v_{1||}^2}\frac{v_{1||}}{v_{CR}}\left(\gamma -1\right)}{\gamma -1 +2/M_{s1}^2}
+\sqrt{\left(\frac{\frac{\gamma +1}{2}+\frac{2}{3}\frac{P_{CR}}{\rho _1v_{1||}^2}\frac{v_{1||}}{v_{CR}}\left(\gamma -1\right)}{\gamma -1 +2/M_{s1}^2}\right)^2-\frac{4}{3}\frac{\frac{P_{CR}}{\rho _1v_{1||}^2}\frac{v_{1||}}{v_{CR}}\left(\gamma -1\right)}{\gamma -1 +2/M_{s1}}}\cr
&\simeq & \frac{\gamma +1+\frac{8}{3}\frac{P_{CR}}{\rho _1v_{1||}^2}\frac{v_{1||}}{v_{CR}}\frac{\gamma -1}{\gamma +1}\left(1-1/M_{S1}^2
\right) + ...}
{\gamma -1 +2/M_{s1}^2}.
\end{eqnarray}
This is the expected result, that energy loss to cosmic rays results in increased shock compression over the usual $\left(\gamma +1\right)\left(\gamma -1
+2/M_s^2\right)$ in terms of the cosmic ray pressure $P_{CR}=n_{CR}p_{CR}v_{CR}=\left(\gamma _{CR} -1\right)E_{CR}$, 
with the cosmic ray energy density and adiabatic index given by $E_{CR}$ and $\gamma _{CR}$ respectively. 
Hence $E_{CR}=3n_{CR}p_{CR}v_{CR}/2$ when the cosmic rays are nonrelativistic and $E_{CR}=3n_{CR}p_{CR}v_{CR}$ when relativistic, as $\gamma _{CR}$ varies between 5/3 and 4/3.
The term in $P_{1w}/\rho _1v_{1||}^2$ complicates things because of the $r-1$ in the denominator, except when $\gamma =2$
and the compression is reduced as $1/M_{s1}^2\rightarrow 1/M_{s1}^2+2P_{1w}/\rho _1v_{1||}^2$. We expect similar behavior at lower $\gamma$, 
because the waves represent a $\gamma =2$ contribution to the pre- and post-shock momentum and energy fluxes, in that $v_A=\sqrt{\gamma P/\rho}
=\sqrt{2\times B^2/8\pi/\rho}$.

For a perpendicular shock ($\theta _1 = 90^{\circ}$), with $P_{1w}\rightarrow 0$ and $v_{1||}<<v_{CR}$ as before,
\begin{equation}
\frac{\gamma +1}{2r}-\frac{\gamma -1}{2}-\frac{1}{M_{s1}^2}
+\frac{2}{3}\frac{n_{CR}p_{CR}}{\rho _1v_{1||}}\left(r-1\right)\left\{\frac{\gamma -1}{r^2}\right\}+\frac{\gamma}{2}\frac{\left(r-1\right)}{M_{A1}^2}=0.
\end{equation}
With $n_{CR}=0$ this reduces to a quadratic equation with solution
\begin{equation}
r=\frac{\gamma +1}{\gamma -1+2/M_{s1}^2+\gamma /M_{A1}^2} + ... =r_0
\end{equation}
The full solution of the cubic equation 9 is very complicated. An easier approach is to write $r=r_0+\delta$ with $\delta << r_0$,
and substitute into equation 9 to find solution
\begin{equation}
r= \frac{\gamma +1+\frac{8}{3}\frac{P_{CR}}{\rho _1v_{1||}^2}\frac{v_{1||}}{v_{CR}}\frac{\gamma -1}{\gamma +1}\left(1-1/M_{S1}^2
-\gamma /2M_{A1}^2\right) + ...}
{\gamma -1 +2/M_{s1}^2 +\gamma /M_{A1}^2},
\end{equation}
which is the same as equation 8 with the substitution $1/M_{S1}^2 \rightarrow 1/M_{s1}^2+\gamma/2M_{A1}^2$.

Before proceeding to an evaluation of the first critical Mach number following \citet{edmiston84} below, we first make some comments about the
presence (or absence) of other critical Mach numbers in the cosmic ray modified shock. We are considering a shock driven by a solar coronal mass ejection
starting off as a waves, gradually steepening to a shock and beginning to accelerate particles. For the relevant shock parameters, the cosmic rays (or
solar energetic particles) remain nonrelativistic with $\gamma _{CR}=5/3$. Thus critical Mach numbers in hydrodynamic shocks associated with multi-valued downstream solutions
that arise with $\gamma _{CR}=4/3$ \citep[e.g.][]{drury81, drury83} do not concern us here. Nonrelativistic cosmic rays of sufficient pressure may still smooth out the discontinuous 
shock transition. \citet{becker01} in their Figure 2a give a range of sonic Mach numbers $M_{g0}=\sqrt{\rho v^2/\gamma _GP_G}$ and $M_{c0}=\sqrt{\rho v^2/\gamma _{CR}P_{CR}}$ for which a discontinuous shock transition is guaranteed. For $\gamma _{CR}=\gamma _G=5/3$, for $M_{g0} > 5.5$,
the discontinuity vanishes, even when the cosmic ray pressure goes to zero. This falls in the range of sonic Mach number ($M_{g0}\simeq M_f\sqrt{2/\gamma\beta +\sin ^2\theta}$ for $\beta << 1$)
relevant to our case, but is based on a model where the
cosmic rays stream with the velocity of the background gas. Appendix B shows that this inconsistency disappears if the cosmic rays are treated as being
trapped in the acceleration process at the shock, up to a maximum cosmic ray pressure given by equation B19 which evaluates to $P_{CR,max}/\rho v^2 = 0.2$ 
for $M\rightarrow \infty$. At lower $M$ relevant to Figure 2 below, $P_{CR,max}/\rho v^2 \sim 0.1$. This becomes relevant for the highest energy case at the highest values of $\beta$
considered. Finally, \citet{vink14} derive critical sonic Mach numbers of $\sqrt{5} - 2.5$
depending on plasma $\beta$ and shock geometry for the existence of a cosmic ray precursor, which also lands in the middle of the range of Mach numbers
considered below. But these results do not apply in a case with pre-existing cosmic rays (where a precursor of some sort is inevitable), and so are not considered further.

\section{First Critical Mach Numbers} \label{sec:mach}
We solve the full equation 6 for the shock compression, $r$, for a range of plasma $\beta =1.2 M_{A1}^2/M_{s1}^2 = 8\pi P/B^2$ for $\gamma =5/3$
between 0 and 1, and for shock obliquities $\theta _1 = 0-90$ degrees. Cases of different SEP particle energy densities are considered, 
with values chosen to match the range of those given by \citet{emslie12}. The upstream wave intensity is then taken from equation 5. Various prescriptions
for $v_{CR}/v_{1||}$ have been tried and we eventually settled on $v_{CR}/v_{1||} = 2/\left(\cos\theta _1 +10^{-6}\right)^2$ chosen to match the injection
energy modeled in \citet{zank06} and to avoid numerical problems at $\cos\theta =0$. It plays very little role in the final results. 
We  then calculate the sonic Mach number in downstream medium,
\begin{eqnarray}
M_{s2}=v_{2||}\sqrt{\frac{\rho _2}{\gamma P_2}}&=&\bigg\{\gamma\left(r-1\right)+\frac{r}{M_{s1}^2}+\frac{r\gamma}{2}\left(\frac{\sin\theta _1}{M_{A1}^2-r\cos ^2\theta _1}\right)^2
\left\{M_{A1}^2\left(1-r^2\right)+2\cos ^2\theta _1\left(r^2-r\right)\right\}\cr
&+& r\gamma\frac{P_{1w}}{\rho _1v_{1||}^2}\left(1-\left(T^2+R^2\right)\right)-\frac{2}{3}\frac{P_{CR}}{\rho _1v_{1||}^2}\left(\frac{v_{1||}}{v_{CR}}
\right)^2\frac{\left(r-1\right)^2}{r}\bigg\}^{-1/2},
\end{eqnarray}
and determine the first critical fast Mach number where $M_{2s}=1$. That this condition corresponds to the first critical Mach number derives from an argument by \citet{coroniti70}, which we briefly summarize here. 
Consider a perpendicular shock where all dissipation is resistive, so that the shock transition occurs on a length scale of order
the magnetic diffusion length. If $M_{s2}>1$, then sound waves coming from the downstream region cannot reach the shock, and it remains stable. 
If $M_{s2}<1$, such sound waves can add kinetic energy to the shock. The shock must either steepen again or begin to reflect incoming ions, since no more 
resistive dissipation is available for the extra kinetic energy. This is the onset of supercriticality.

\begin{figure}[t]
\centerline{\includegraphics[width=3.in]{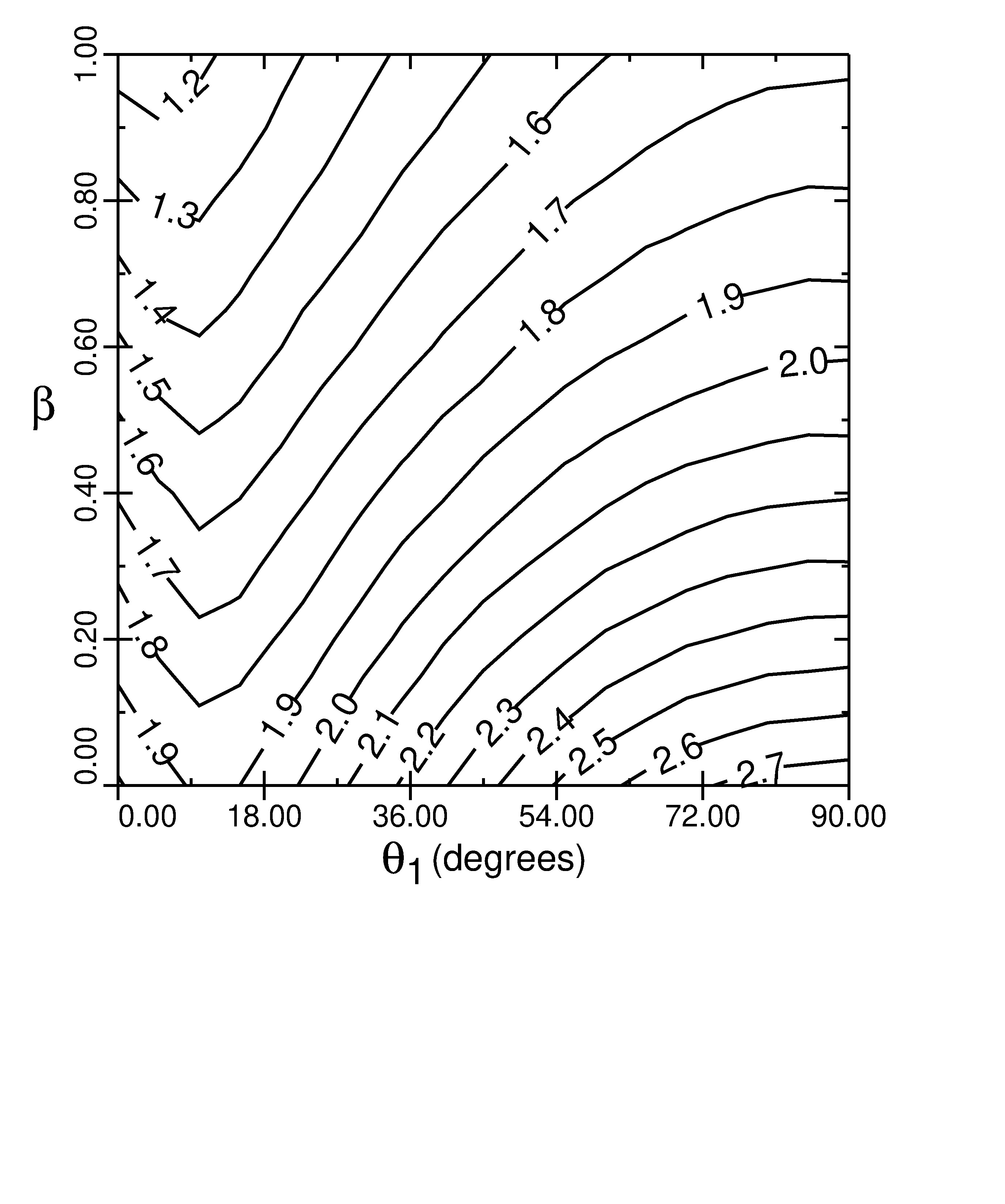}\includegraphics[width=3.in]{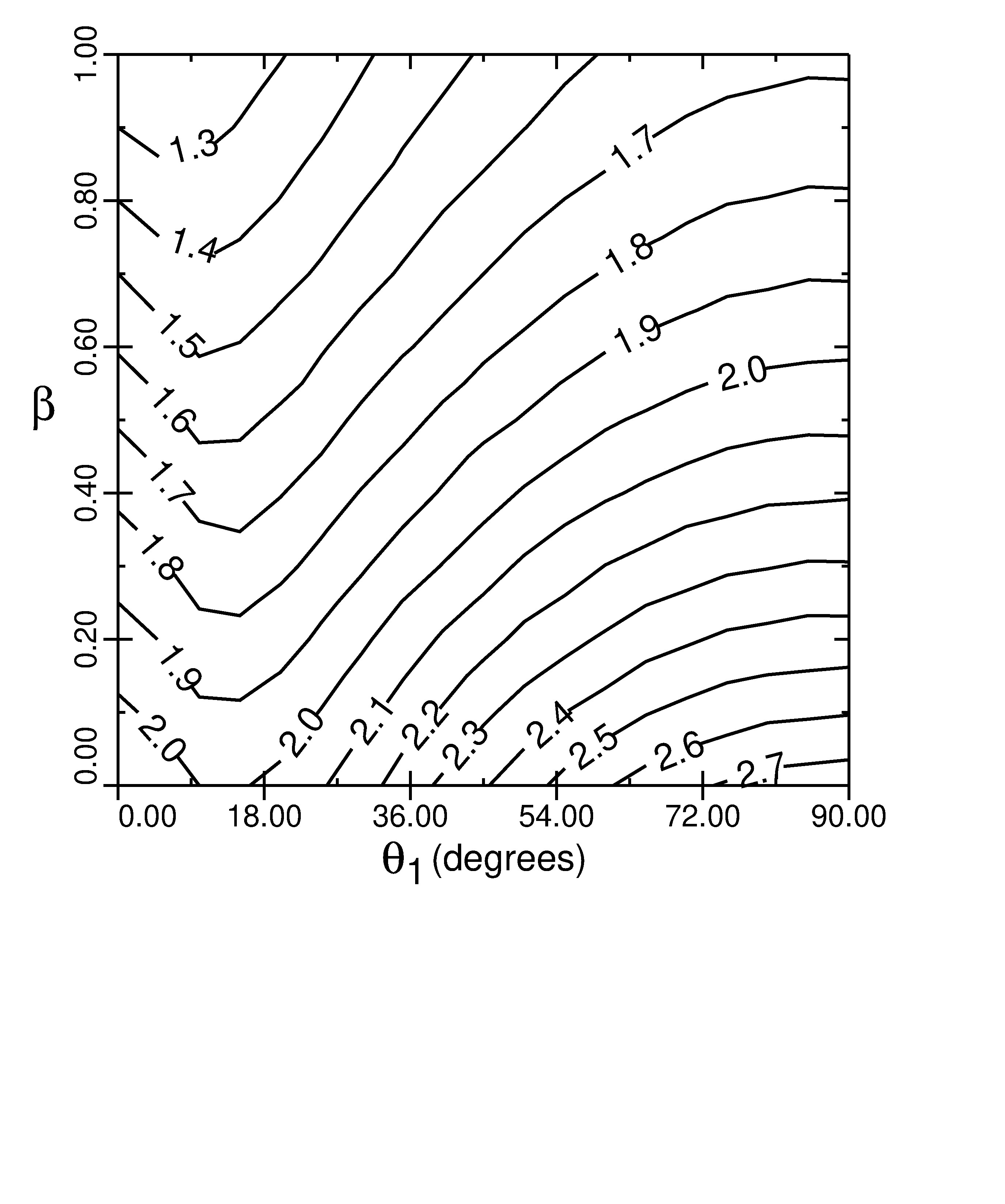}}
\vspace{-0.9truein}
\centerline{\includegraphics[width=3.in]{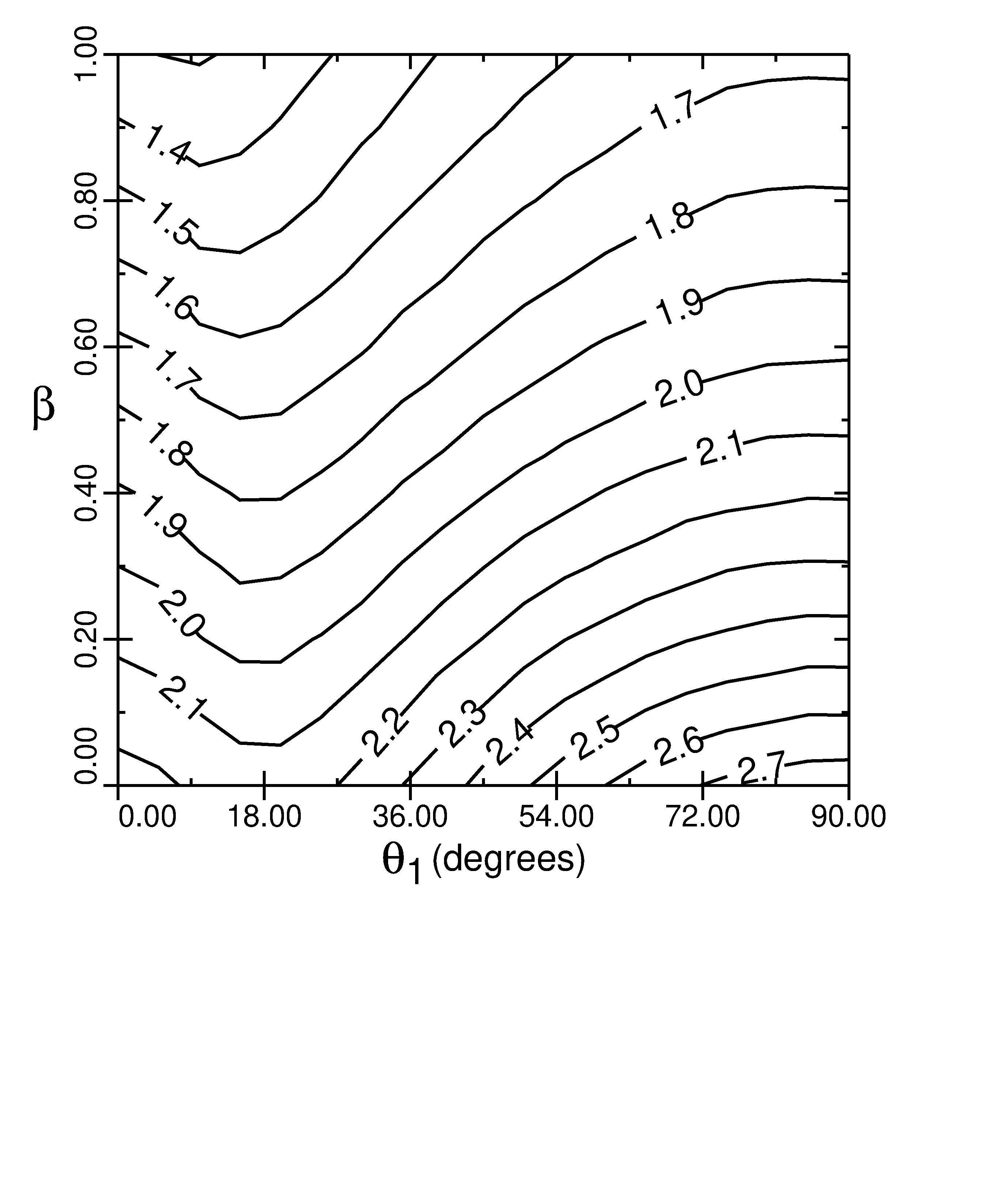}\includegraphics[width=3.in]{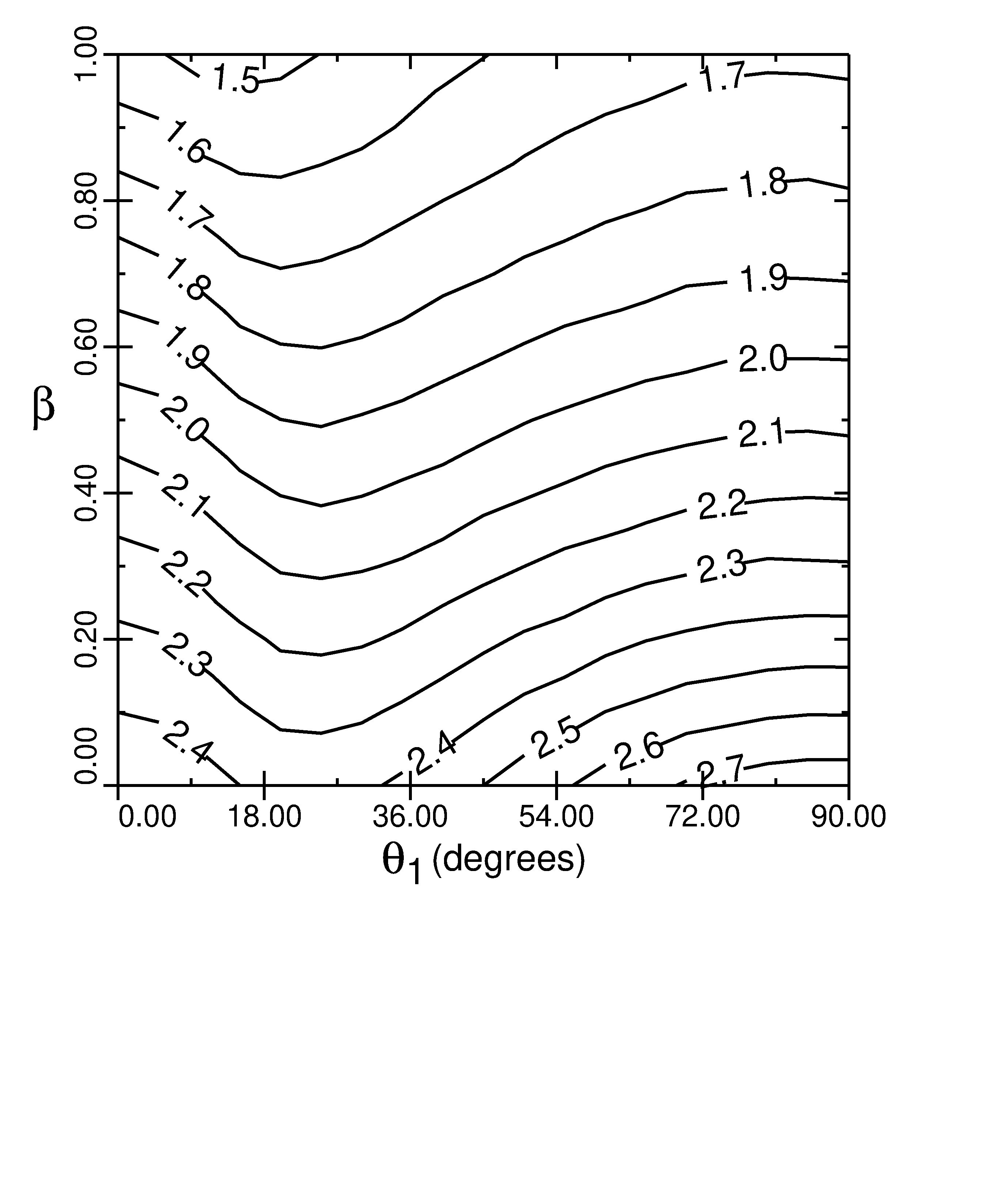}}
\vspace{-0.9truein}
\caption{Contour plots of the first critical fast Mach number in the plasma $\beta$ - $\theta _1$ plane for $E_{CR}/\rho _1v_{1||}^2 = 0.025$, 0.05, 
0.1 and 0.2 in top left, top right, bottom left, and bottom right respectively. \label{critical1}}
\end{figure}

Figure \ref{critical1} plots the first critical fast Mach number for values of (nonrelativistic)
 $E_{CR}/\rho _1v_{1||}^2 = 0.025$, 0.05, 0.1 and 0.2 in the $\beta - \theta _1$ plane, for comparison with Figure 1 right panel which gives the
case of $P_{CR}=0$ from \citet{edmiston84}. The terms involving cosmic rays and turbulence increase the first critical Mach number for 
quasi-parallel shocks. This effect is due to the waves accompanying the cosmic rays, which diminish in intensity as the shock becomes more oblique.
Critical Mach numbers for perpendicular shocks are almost unchanged from the $P_{CR}=0$ case, and at the highest cosmic ray pressure
considered, $E_{CR}/\rho _1v_{1||}^2=0.2$, the critical Mach number is almost independent of shock obliquity. The increase in critical Mach
number can be understood in terms of the extra dissipation available to the shock. The pre-existing waves are strongly amplified in intensity
on passing through the shock, and this represents an extra mode of dissipation available to the shock beyond the resistivity, thereby increasing the
Mach number at which the shock becomes supercritical. Note that this shock amplification of pre-existing waves is distinct from ion reflection and the
waves generated by this at the onset of supercriticality. 

\begin{figure}[t]
\centerline{\includegraphics[width=3.in]{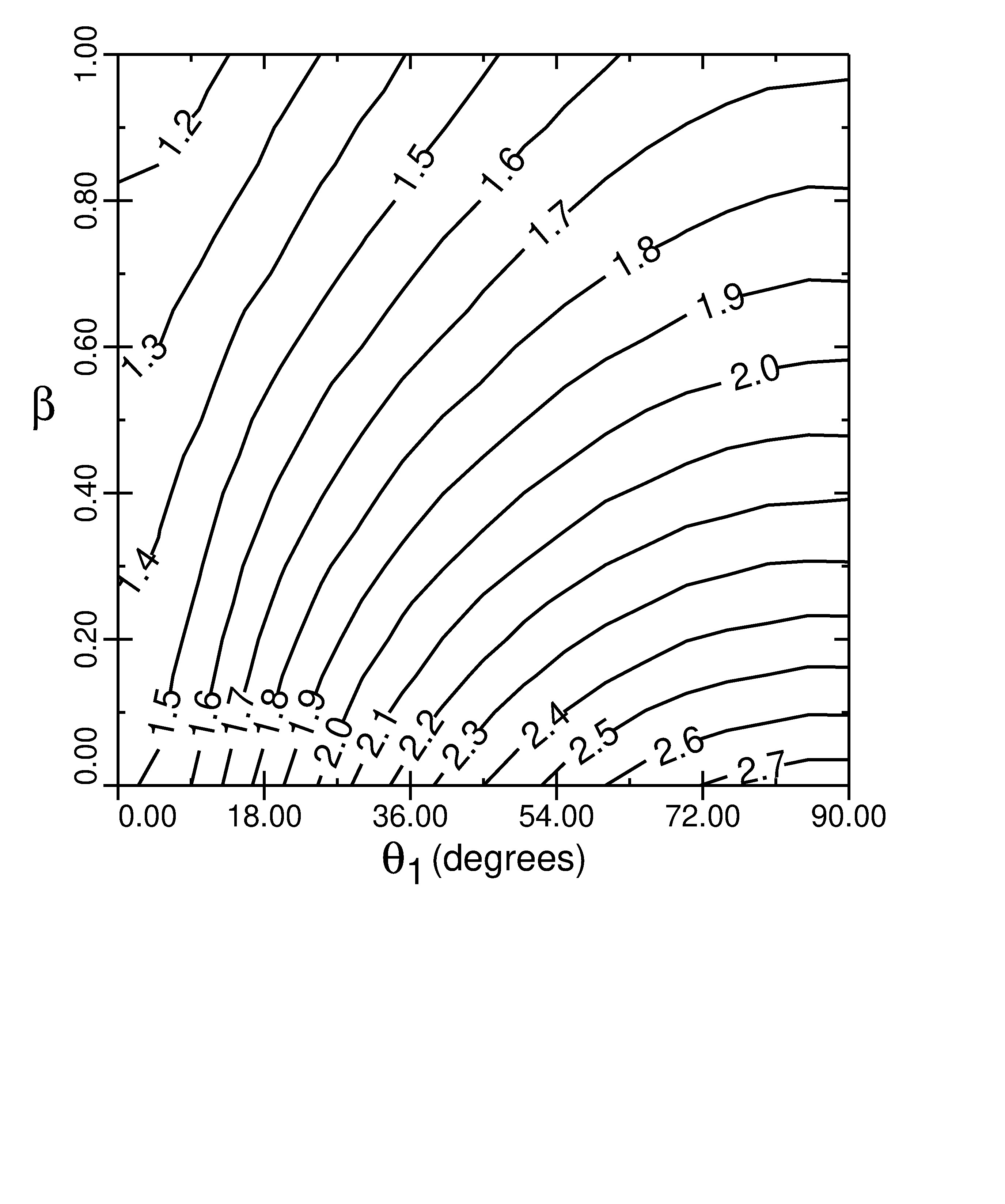}\includegraphics[width=3.in]{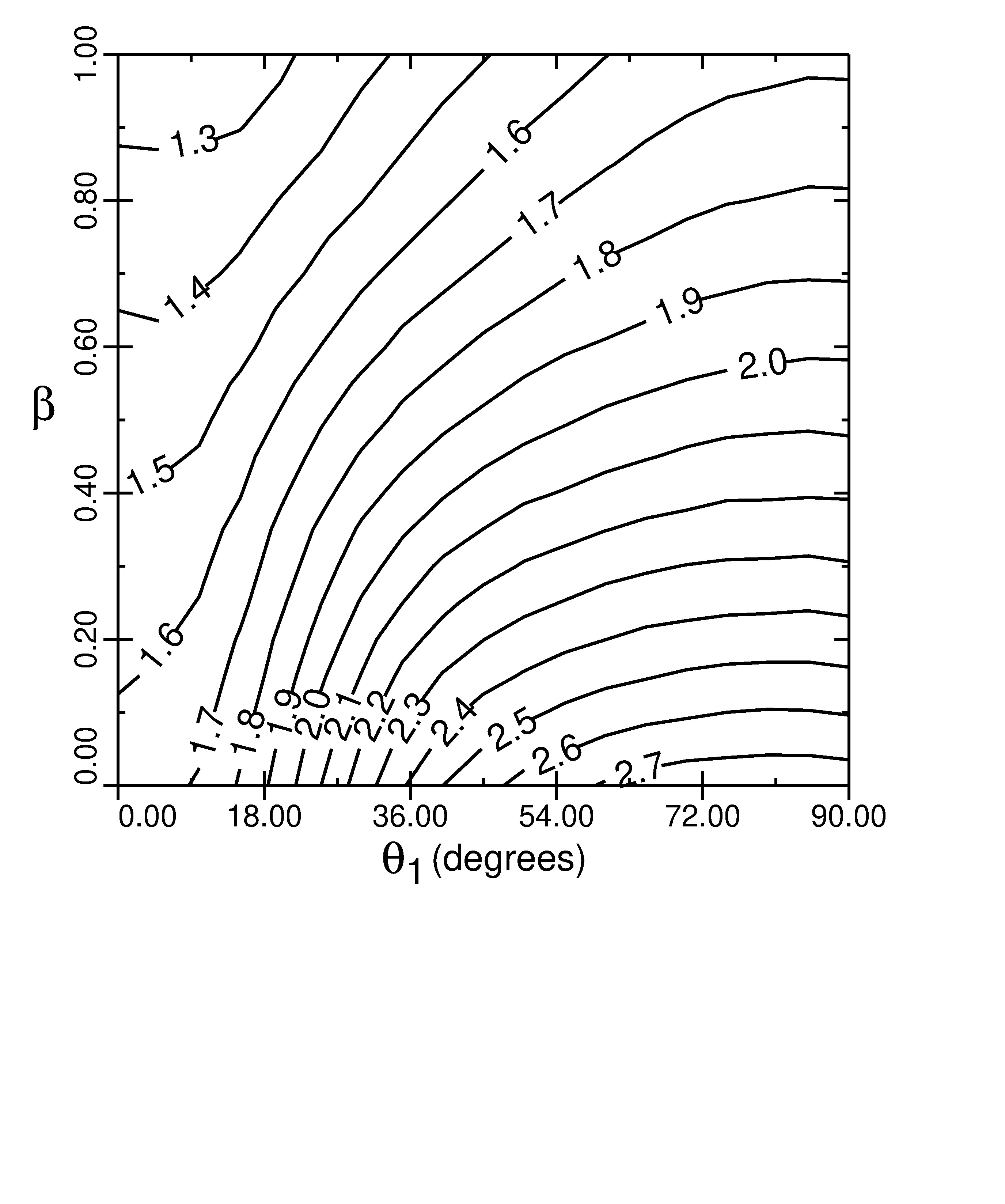}}
\vspace{-0.9truein}
\centerline{\includegraphics[width=3.in]{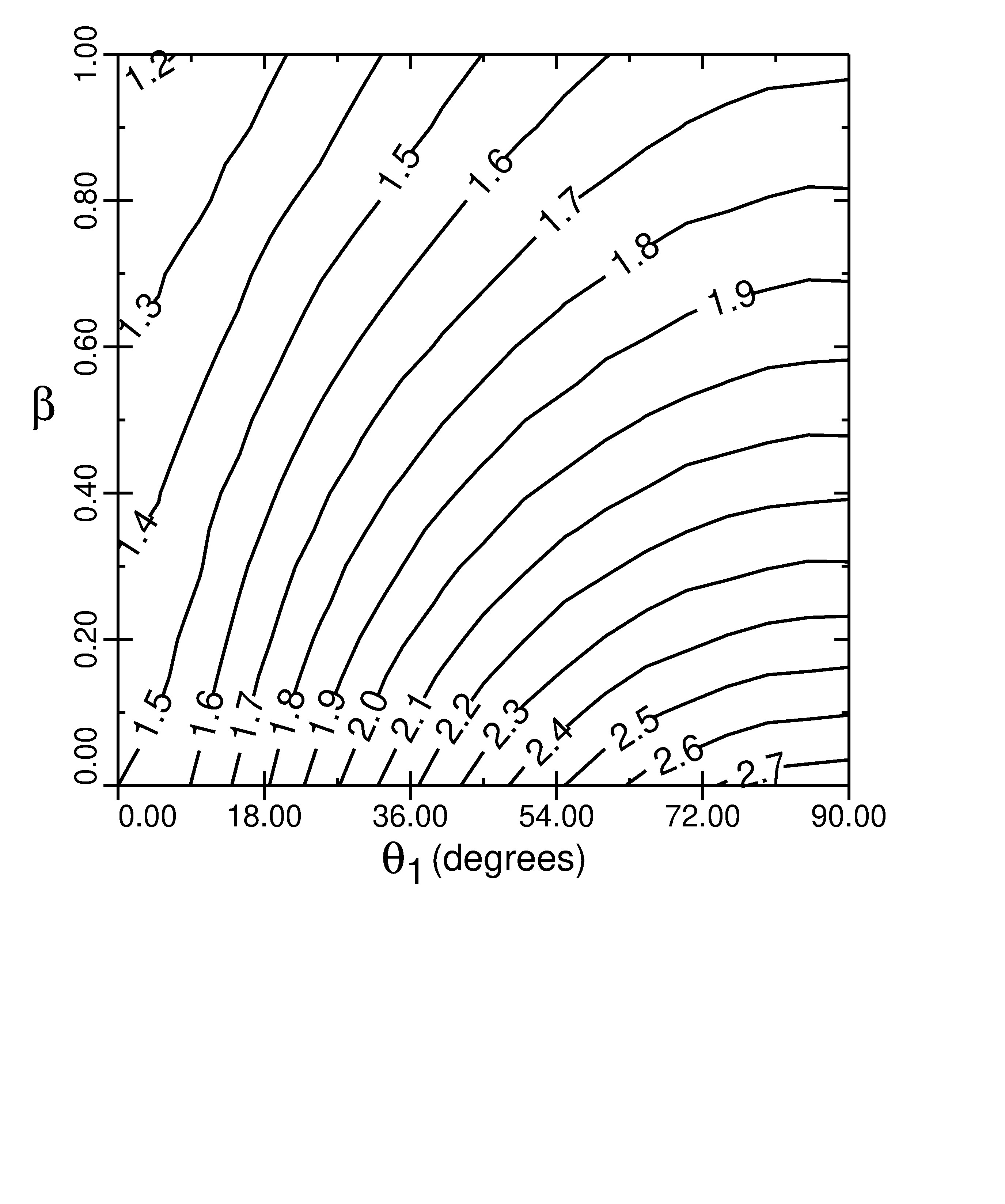}\includegraphics[width=3.in]{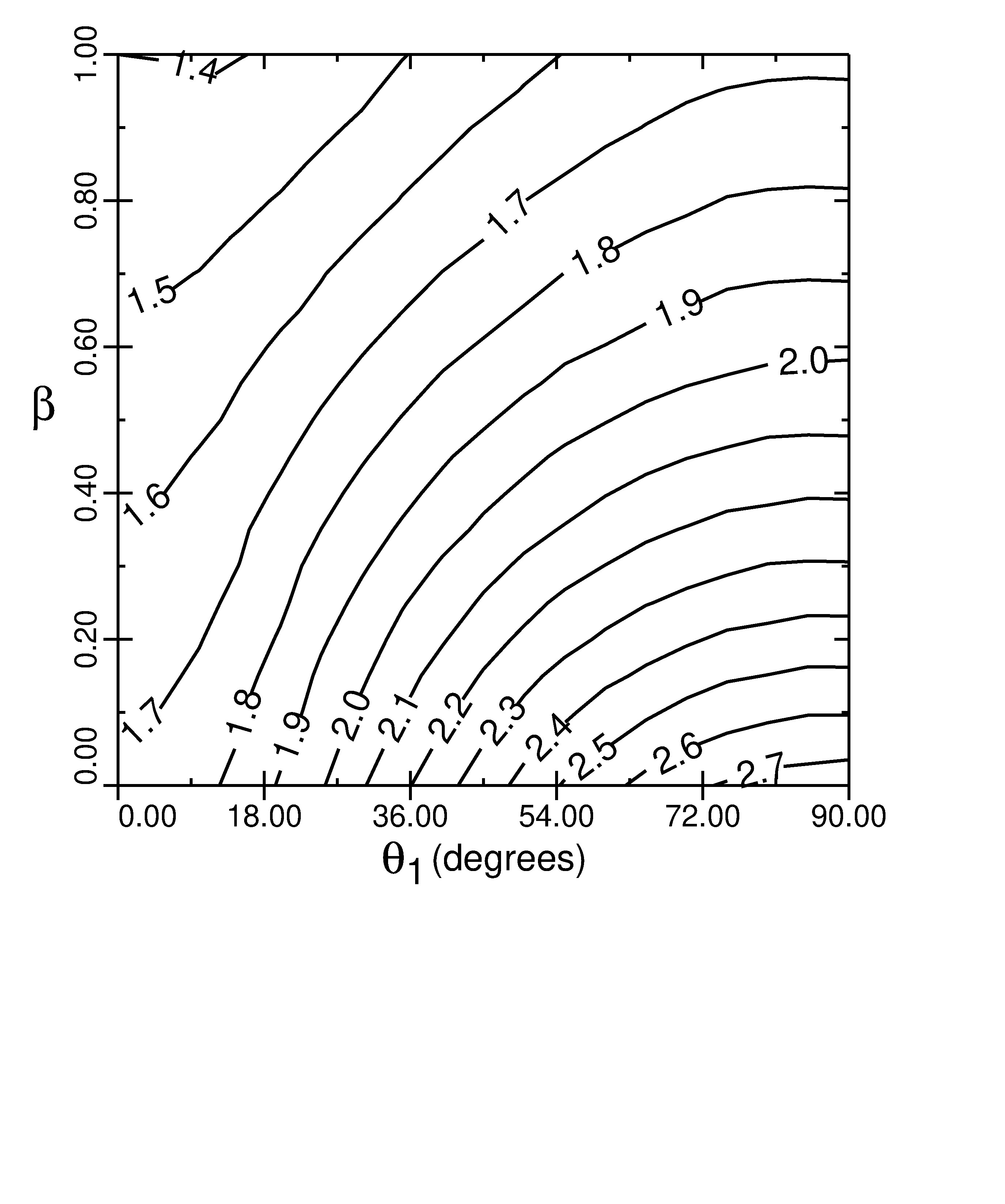}}
\vspace{-0.9truein}
\centerline{\includegraphics[width=3.in]{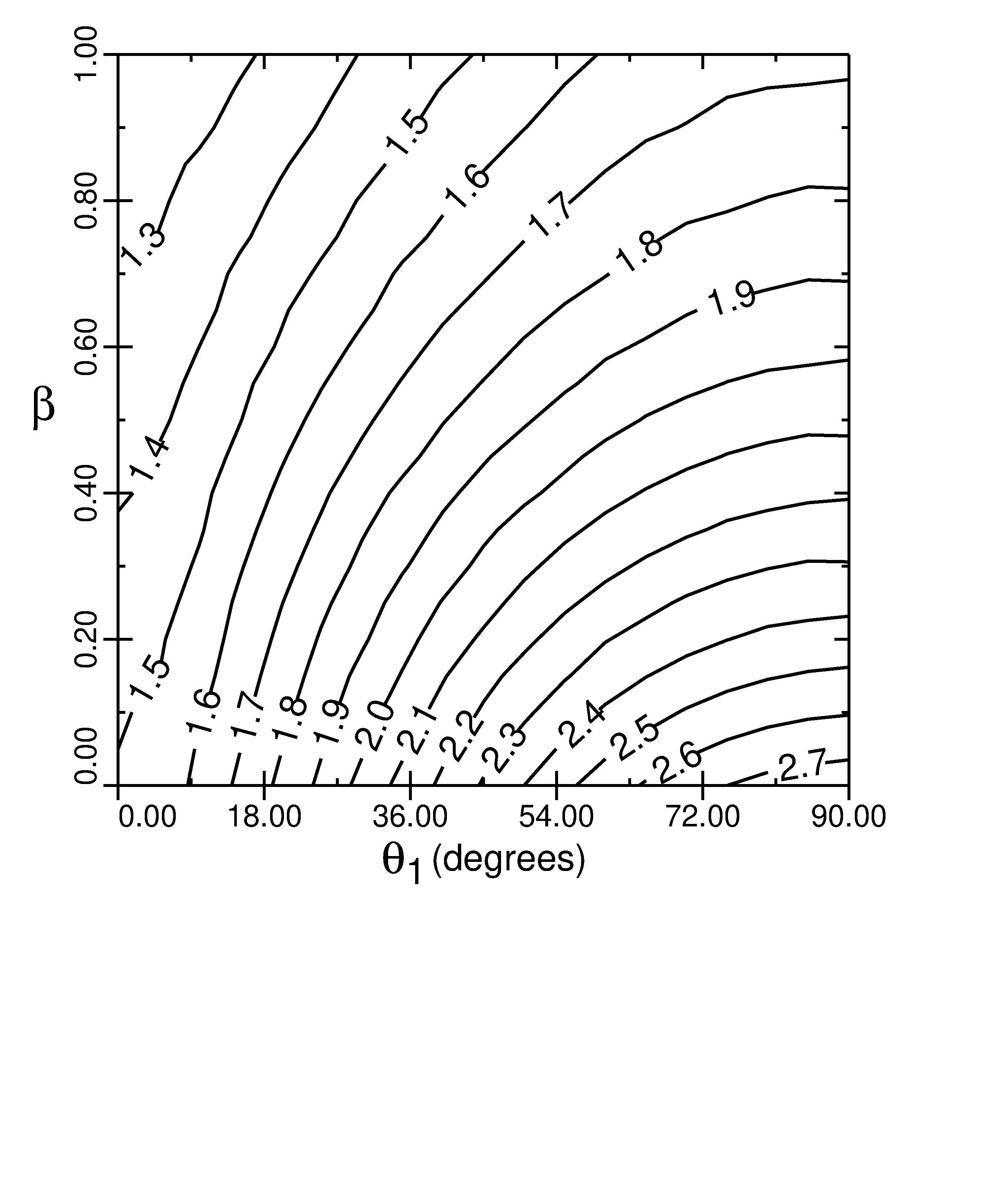}\includegraphics[width=3.in]{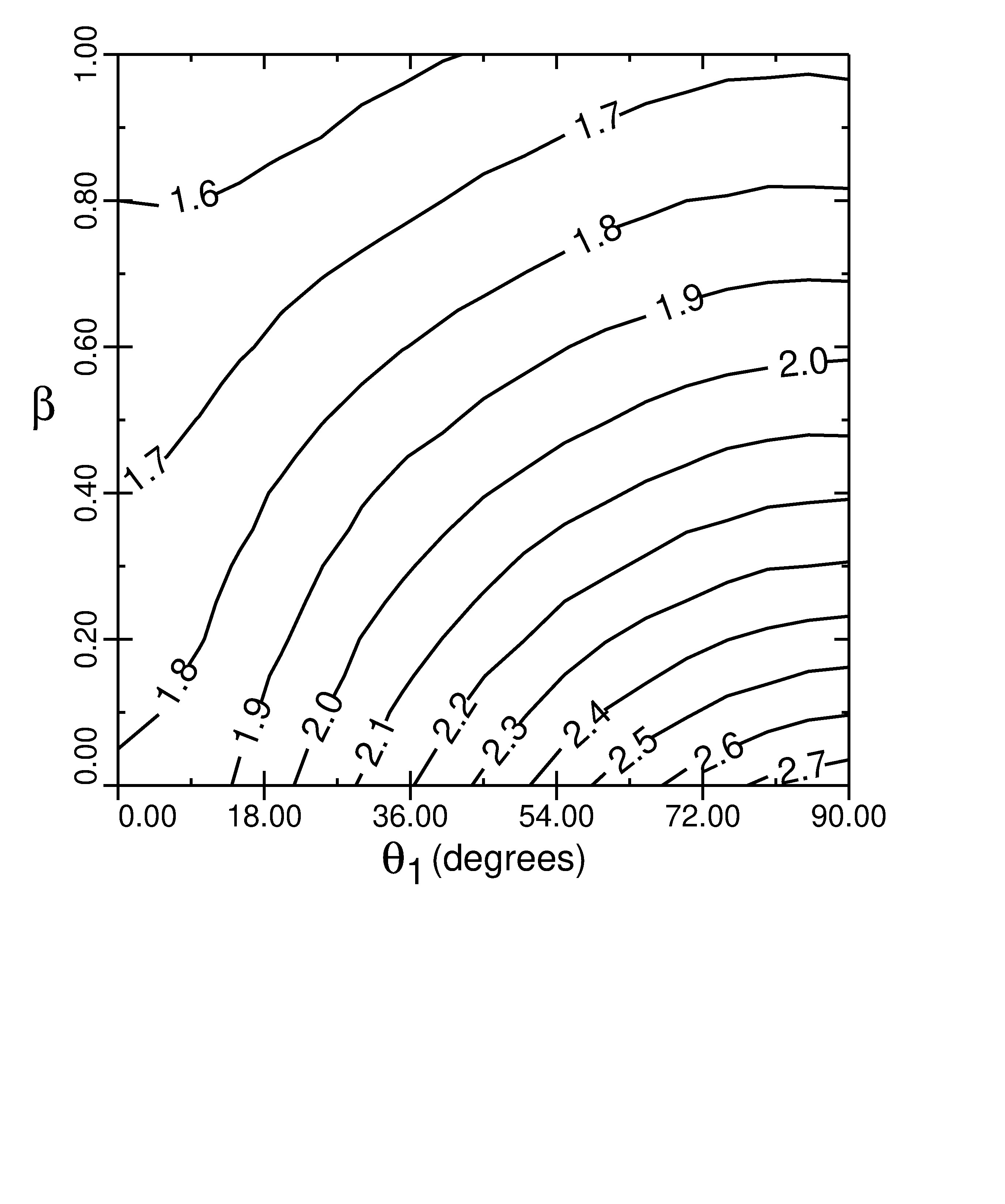}}
\vspace{-0.9truein}
\caption{Contour plots of the first critical fast Mach number in the plasma $\beta$ - $\theta _1$ plane for $E_{CR}/\rho _1v_{1||}^2 = 0.025$ (left)
and 0.1 (right) panels, showing initially forward propagating waves only (top), initially balanced waves (middle) and initially backward
propagating waves only (bottom).\label{critical2}}
\end{figure}

Equation 4 assumed equal wave intensities propagating forward and backward in the upstream and downstream media. Relaxing this approximation, following equation 3, we write in the energy equation
\begin{eqnarray}
P_{1w}v_{1||}&\rightarrow &P_{1wf}\left(v_{1||}+v_{A1}\cos\theta _1\right) + P_{1wb}\left(v_{1||}-v_{A1}\cos\theta _1\right)=P_{1w}v_{1||}
+\left(P_{1wf}-P_{1wb}\right)v_{A1}\cos\theta _1\cr
P_{2w}v_{2||}&\rightarrow &\left\{\sum _{pol}\frac{P_{1wf}}{2}\left(T_f^2+R_f^2\right)+\sum _{pol}\frac{P_{1wb}}{2}
\left(T_b^2+R_b^2\right)\right\}v_{2||}+
\left\{\sum _{pol}\frac{P_{1wf}}{2}\left(T_f^2-R_f^2\right)-\sum _{pol}\frac{P_{1wb}}{2}\left(T_b^2-R_b^2\right)\right\}v_{A2}\cos\theta _{2w}.
\end{eqnarray}
With the new terms in $v_{A1}\cos\theta _1$ and $v_{A2}\cos\theta _2$ equation 5 becomes
\begin{eqnarray}
\bigg\{\frac{a}{r}&-&\frac{1}{M_{s1}^2}+\frac{P_{1w}}{\rho _1v_{1||}^2}\frac{\gamma -2r\left(\gamma -1\right)
-\sum _{pol}\left(T_b^2+T_f^2+R_b^2+R_f^2\right)\left(2-\gamma\right)/4}{r-1}\cr
&+&2r\frac{\gamma -1}{r-1}\frac{P_{1w}}{\rho _1v_{1||}^2}\left\{\frac{P_{1wb}-P_{1wf}}{P_{1w}}\frac{\cos\theta _1}{M_{A1}}+
\left[\sum _{pol}\frac{P_{1wf}}{2P_{1w}}\left(T_f^2-R_f^2\right)-
\sum _{pol}\frac{P_{1wb}}{2P_{1w}}\left(T_b^2-R_b^2\right)\right]\frac{v_{A2}}{v_{2||}}\cos\theta _{2w}\right\}\cr
&+&\frac{F_1/r-F_2/r^2}{\rho _1v_{1||}}\frac{\gamma}{\gamma -1}-\frac{E_1-E_2/r}{\rho_1v_{1||}^2}\frac{\gamma -1}{r-1}\bigg\}\left(\frac{1}{r}
-\frac{\cos ^2\theta _1}{M_{A1}^2}\right)^2 \
- \frac{\sin ^2\theta_1}{M_{A1}^2r}\left\{\frac{1}{r}\left(a-\frac{1-r}{2}\right)-\frac{a\cos ^2\theta_1}{M_{A1}^2}\right\}=0.
\end{eqnarray}
We plot in Figure \ref{critical2} the Critical Mach numbers as before for the cases of $E_{CR}/\rho _1v_{1||}^2= 0.025$ and 0.1 for the three cases
$P_{1wf}/P_{1w}=1$, $P_{1wb}/P_{1w}=0$, (top panels), $P_{1wf}/P_{1w}=0.5$, $P_{1wb}/P_{1w}=0.5$, (middle panels), and
$P_{1wf}/P_{1w}=0$, $P_{1wb}/P_{1w}=1$, (bottom panels). In both cases, the Critical Mach numbers are highest for the last, with waves initially propagating away from the shock, and lowest with initially forward propagating waves. This is because the initially backward propagating waves accumulate
at the shock more than the forward propagating waves, providing more dissipation by their amplification. The case with initially balanced waves is 
intermediate, and all cases presented here have lower Critical Mach numbers than the case presented above, with balanced waves in both upstream and
downstream. The reason is that initially balanced waves upstream $P_{1wf}=P_{1wb}$ {\em do not} automatically lead to balanced waves
downstream ($P_{2wf}=P_{2wb}$). In the calculation above it has been tacitly assumed the shock has expended yet more energy in isotropizing the 
waves, over and above that required to amplify them, and this extra dissipation leads to higher critical Mach numbers than in the middle panels of 
Figure \ref{critical2}.


\section{Discussion and Conclusions}
The main new effect in this paper has been the amplification of MHD waves passing through a shock. We have calculated wave refraction, and
transmission and reflection coefficients according to geometric optics. By way of contrast, other works \citep{zank12,zank17,wang22a} treat the transport of
incompressible and nearly incompressible turbulence in inhomogeneous flows more generally, without refering to specific wave properties of the fluctuations. \citet{zank21}, in a treatment more similar to ours, considers
the transmission of various perturbations with magnetic field and velocity components in the plane of a perpendicular shock. Thus at high plasma beta, oscillating acoustic modes, and zero frequency entropy, vortical and magnetic island modes are found to be amplified in amplitude by factors of order 10, similar
to what we find for magnetosonic waves at low plasma beta in this paper. \citet{nakanotani22} perform a 2D hybrid kinetic simulation with a fully turbulent
upstream medium and find similar results, together with diffusive shock acceleration at the shock front itself. \citet{nakanotani21} use similar
methods to consider a different particle acceleration mechanism, where it occurs as magnetic islands
merge and contract postshock. The origin of the turbulent structures is not discussed, but the energetic particle signature resembles that observed by
Voyager at the heliospheric termination shock.

Another extreme lies in shocks not just modified by the accelerated particles, but actually {\em mediated} by accelerated particles. \citet{mostafavi18} find the
heliospheric termination shock to be formed and dominated by processes in the pickup ions rather than the (quasi-)thermal plasma. \citet{wang22b} consider
a similar problem with application to supernova remnant shock waves in the interstellar medium, using a coupled system of equations to describe the
gas, cosmic rays and turbulence, based in part onworks cited above. Both sets of authors find the shock discontinuity disappears, and the transition
between upstream and downstream is broadened.

The amplification of waves passing through an MHD shock represents an extra dissipation mechanism for kinetic energy entering the shock. 
This increases the first critical fast Mach number,
which delays the onset of Type II and Type III radio bursts, and possibly also the release of SEPs  \citep{zhu18}, which
would accentuate the variability of SEP events. The existence of a seed particle 
population that can be accelerated by a sub-critical shock not only affects the injection process, but the waves they generate affect how long the shock may accelerate particles before releasing them,
and hence their eventual energy and spectrum. The inference of \citet{zhu18} also suggests that Type II radio bursts (and also Type III but this should be
obvious) require accelerated particle {\em release} from the shock. The Type II radio burst does not result from the usual cosmic ray precursor 
disturbing the upstream medium ahead of the shock, but from particles that can escape this precursor and move further upstream. The growth
of Langmuir waves driven by escaping cosmic rays would be faster than that due to the quasi-isotropic cosmic ray distribution drifting with the shock,
and the higher Langmuir wave energy density would allow faster conversion to electro-magnetic waves before being overrun by the shock. This
type of instability is qualitatively different to those discussed by \citet{rakowski08} and \citet{laming14}, where it is the quasi-isotropic
cosmic ray precursor drifting with the shock that excites electrostatic waves in the precursor that damp by heating electrons.
The usual stated criterion for cosmic ray escape from the shock acceleration process is that the particle gyroradius must be larger than the 
characteristic shock dimension. This plausibly
comes about when the shock becomes supercritical, because ions incident from upstream are now reflected back to the upstream by the
cross shock potential. This is not constant, but varies according to a shock reformation cycle, and periods of low cross shock potential and hence
low magnetic field jump will allow cosmic ray escape. 

The idea that pre-existing suprathermal seed particles can be accelerated at a sub-critical shock and then begin to be released when the shock
becomes supercritical requires some reinterpretation of \citet{laming13}. This paper calculates the growth rate of waves, assumed upstream of a
shock, by a distribution of seed particles that have either reflected from, or are drifting with, the same shock. Such a scenario tacitly assumes a super-critical
shock. In the case of a sub-critical shock, the seed particles will not automatically reflect from the shock front, but will convect into the downstream region.
There they may excite waves, isotropize themselves and ultimately drift back to the upstream, where the same processes would start the 
diffusive shock acceleration
cycle. Thus the calculations are still relevant, but the obliquity needs to be taken with respect to the {\em downstream} magnetic field of a laminar
sub-critical shock, which is moving
with respect to the seed particle distribution with the postshock flow speed given in terms of the shock velocity $v_s$ and compression $r$ by $v_s\left(1-1/r\right)$.

In this paper we have investigated the effect on shock criticality of a population of pre-existing seed particles and associated waves in the upstream 
medium into which the shock is propagating. The accelerated particles themselves do not have much effect, but the associated waves can be strongly
amplified on passage through the shock and this represents an alternative dissipation mechanism for the shock kinetic energy, beyond that provided by
the plasma resistivity. This extra dissipation increases the Mach numbers at which the shock can remain laminar, and the first critical fast Mach number
at which the shock has previously been assumed to commence particle acceleration is increased. However we argue based on observations of Type II and
III radio bursts and the careful study of \citet{zhu18} that these Mach numbers should be more properly associated the beginning of the {\em release} 
of energetic particles from the shock, and that given the right conditions, particle acceleration can commence as soon as the shock forms and
thermal plasma becomes compressed. Thus given an initial seed particle distribution, the path to SEP acceleration and release close to the Sun
might be more specified in terms of basic physics than has previously been appreciated, and it will be a goal of future work to investigate this in more detail.

\begin{acknowledgments}
This work was supported by basic research funds of the Office of Naval Research, and by the NASA Heliophysics Theory, Modeling and Simulation program
grant 80HQTR20T0067.
\end{acknowledgments}

\appendix

\section{Full Transmission and Reflection Coefficients}
We first collect some standard results for oblique shocks \citep[e.g.][]{achterberg86,laming15,melrose86}. The downstream magnetic field is
\begin{equation}
B_2 = rB_1\frac{\left(M_{A1}^2-\cos ^2\theta _1\right)}{\left(M_{A1}^2-r\cos ^2\theta _1\right)}\frac{\sin\theta _1}{\sin\theta _2}=B_1\frac{\cos\theta _1}
{\cos\theta _2},
\end{equation}
while the downstream perpendicular flow velocity component is
\begin{equation}
v_{2\perp}=v_1\frac{\left(r-1\right)\sin\theta _1\cos\theta _1}{M_{A1}^2-r\cos ^2\theta _1}.
\end{equation}
The tangent of the angle, $\theta _2$, between the shock normal and the downstream magnetic field is
\begin{equation}
\tan\theta _2=r\tan\theta _1+r\frac{\left(r-1\right)\sin\theta _1\cos\theta _1}{M_{A1}^2-r\cos ^2\theta _1}=
\frac{r\tan\theta _1}{M_{A1}^2-r\cos ^2\theta _1}\left(M_{A1}^2-\cos ^2\theta _1\right),
\end{equation}
which corrects a typographical error in equation 39 in \citet{laming15}.
This angle $\theta _2$ is {\em different} to the angles at which initially parallel propagating waves in the upstream region 1 are
refracted to in region 2. For Alfv\'en waves, or the Alfv\'en polarization component of an upstream circularly polarized waves, the angle between the
propagation direction and the shock normal in the downstream region, $\theta _{2w}$, is 
\begin{eqnarray}
\cot\theta _{2w}&=&\frac{M_{A1}r\cot\theta _1+r^{3/2}\sin\theta _1-r/\sin\theta _1}{M_{A1}-r^{1/2}\cos\theta _1}- r\frac{\left(r-1\right)\sin\theta _1\cos\theta _1}{M_{A1}^2-r\cos ^2\theta _1}\cr
&=&\frac{r\cot\theta _1}{M_{A1}^2-r\cos ^2\theta _1}\left(M_{A1}^2+\frac{M_{A1}}{\cos\theta _1}\left(r^{1/2}-1\right)
-r^{1/2} +\sin ^2\theta _1\right).
\end{eqnarray}
This corrects equation 11 in \citet{laming15}, which was missing a term $r^{1/2}\cos\theta _1\tan\alpha/M_{A1}$ on the right hand side.
As well as being different to $\theta _2$, $\theta _{2w}$ is also different to the angle at which initially parallel propagating magnetosonic waves, or the magnetosonic
polarization component of a circularly polarized waves, will refract. For postshock magnetosonic waves at low plasma $\beta$ 
\citep[][equation 5 with $V_{A1}/V_{A2}=\sin\theta _1/\sin\theta _{2w}$]{laming15}
\begin{equation}
\cot\theta _{2w}=r\cot\theta _1-r\frac{\left(r-1\right)\sin\theta _1\cos\theta _1}{M_{A1}^2-r\cos ^2\theta _1}
= \frac{r\cot\theta _1}{M_{A1}^2-r\cos ^2\theta _1}\left(M_{A1}^2-r +\sin ^2\theta _1\right)
\end{equation}
As $M_{A1}\rightarrow \infty$, $\theta _{2w}$ is the same for both Alfv\'en and magnetosonic waves, $\theta _{2w}=\cot ^{-1}\left(r\cot\theta _1\right)$, 
but remains different to $\theta _2$. At high plasma $\beta$, less relevant in this work, magnetosonic waves refract more like Alfv\'en waves.

The transmission and reflection coefficients for the Alfv\'en polarization are relatively simple, and do not depend on $\theta _{2w}$ since the 
Alfv\'en polarization has $\delta {\rm B}$ orthogonal to the plane of incidence. They are \citep[][equation 15]{laming15}
\begin{equation}
T_f=\frac{1}{2}\frac{\left(M_{A1}+\cos\theta _1\right)\left(r^{1/2}+1\right)r^{1/2}}{M_{A1}+r^{1/2}\cos\theta _1},
\end{equation}
\begin{equation}
R_f=\frac{1}{2}\frac{\left(M_{A1}+\cos\theta _1\right)\left(r^{1/2}-1\right)r^{1/2}}{M_{A1}-r^{1/2}\cos\theta _1},
\end{equation}
\begin{equation}
T_b=\frac{1}{2}\frac{\left(M_{A1}-\cos\theta _1\right)\left(r^{1/2}+1\right)r^{1/2}}{M_{A1}-r^{1/2}\cos\theta _1},
\end{equation}
and
\begin{equation}
R_b=\frac{1}{2}\frac{\left(M_{A1}-\cos\theta _1\right)\left(r^{1/2}-1\right)r^{1/2}}{M_{A1}+r^{1/2}\cos\theta _1}.
\end{equation}

Equivalent expressions for the magnetosonic polarization are much more involved. As in \citet{laming15}, who only gave expressions for the
limit $M_A\rightarrow\infty$, we neglect the effect of the passing wave on the motion of the shock front itself, but otherwise retain all terms
to derive the more general expressions; 
\begin{eqnarray}
T_f&=&\frac{1}{2}\frac{\left(M_{A1}\cos\theta _1+1\right)r^{1/2}\left\{\left(B_2/B_1\right)\cos\left(\theta _2+\theta _{2w}\right)+
M_{A1}r^{1/2}\left(v_{2\perp}/v_{1||}\right)\sin\theta _{2w} -M_{A1}r^{-1/2}\cos\theta _{2w}\right\} }
{\left(B_2/B_1\right)\cos\left(\theta _2+\theta _{2w}\right)\cos\left(\theta _2-\theta _{2w}\right)+
M_{A1}^2r\left(v_{2\perp}/v_{1||}\right)^2\sin ^2\theta _{2w} -\left(M_{A1}^2/r\right)\cos ^2\theta _{2w}}\cr
&+&\frac{1}{2}\frac{\left(M_{A1}\cos\theta _1+\cos 2\theta _1\right)\left\{\left(B_2/B_1\right)\cos\left(\theta _2-\theta _{2w}\right) -
M_{A1}r^{1/2}\left(v_{2\perp}/v_{1||}\right)\sin\theta _{2w} -M_{A1}r^{-1/2}\cos\theta _{2w}\right\} }
{\left(B_2/B_1\right)\cos\left(\theta _2+\theta _{2w}\right)\cos\left(\theta _2-\theta _{2w}\right)+
M_{A1}^2r\left(v_{2\perp}/v_{1||}\right)^2\sin ^2\theta _{2w} -\left(M_{A1}^2/r\right)\cos ^2\theta _{2w}},
\end{eqnarray}
\begin{eqnarray}
R_f&=&\frac{1}{2}\frac{\left(M_{A1}\cos\theta _1+1\right)r^{1/2}\left\{-\left(B_2/B_1\right)\cos\left(\theta _2+\theta _{2w}\right)+
M_{A1}r^{1/2}\left(v_{2\perp}/v_{1||}\right)\sin\theta _{2w} -M_{A1}r^{-1/2}\cos\theta _{2w}\right\} }
{\left(B_2/B_1\right)\cos\left(\theta _2+\theta _{2w}\right)\cos\left(\theta _2-\theta _{2w}\right)+
M_{A1}^2r\left(v_{2\perp}/v_{1||}\right)^2\sin ^2\theta _{2w} -\left(M_{A1}^2/r\right)\cos ^2\theta _{2w}}\cr
&+&\frac{1}{2}\frac{\left(M_{A1}\cos\theta _1+\cos 2\theta _1\right)\left\{\left(B_2/B_1\right)\cos\left(\theta _2-\theta _{2w}\right) +
M_{A1}r^{1/2}\left(v_{2\perp}/v_{1||}\right)\sin\theta _{2w} +M_{A1}r^{-1/2}\cos\theta _{2w}\right\} }
{\left(B_2/B_1\right)\cos\left(\theta _2+\theta _{2w}\right)\cos\left(\theta _2-\theta _{2w}\right)+
M_{A1}^2r\left(v_{2\perp}/v_{1||}\right)^2\sin ^2\theta _{2w} -\left(M_{A1}^2/r\right)\cos ^2\theta _{2w}},
\end{eqnarray}
\begin{eqnarray}
T_b&=&\frac{1}{2}\frac{\left(M_{A1}\cos\theta _1-1\right)r^{1/2}\left\{-\left(B_2/B_1\right)\cos\left(\theta _2+\theta _{2w}\right)+
M_{A1}r^{1/2}\left(v_{2\perp}/v_{1||}\right)\sin\theta _{2w} -M_{A1}r^{-1/2}\cos\theta _{2w}\right\} }
{\left(B_2/B_1\right)\cos\left(\theta _2+\theta _{2w}\right)\cos\left(\theta _2-\theta _{2w}\right)+
M_{A1}^2r\left(v_{2\perp}/v_{1||}\right)^2\sin ^2\theta _{2w} -\left(M_{A1}^2/r\right)\cos ^2\theta _{2w}}\cr
&+&\frac{1}{2}\frac{\left(-M_{A1}\cos\theta _1+\cos 2\theta _1\right)\left\{\left(B_2/B_1\right)\cos\left(\theta _2-\theta _{2w}\right) +
M_{A1}r^{1/2}\left(v_{2\perp}/v_{1||}\right)\sin\theta _{2w} +M_{A1}r^{-1/2}\cos\theta _{2w}\right\} }
{\left(B_2/B_1\right)\cos\left(\theta _2+\theta _{2w}\right)\cos\left(\theta _2-\theta _{2w}\right)+
M_{A1}^2r\left(v_{2\perp}/v_{1||}\right)^2\sin ^2\theta _{2w} -\left(M_{A1}^2/r\right)\cos ^2\theta _{2w}},
\end{eqnarray}
and
\begin{eqnarray}
R_b&=&\frac{1}{2}\frac{\left(M_{A1}\cos\theta _1-1\right)r^{1/2}\left\{\left(B_2/B_1\right)\cos\left(\theta _2+\theta _{2w}\right)+
M_{A1}r^{1/2}\left(v_{2\perp}/v_{1||}\right)\sin\theta _{2w} -M_{A1}r^{-1/2}\cos\theta _{2w}\right\} }
{\left(B_2/B_1\right)\cos\left(\theta _2+\theta _{2w}\right)\cos\left(\theta _2-\theta _{2w}\right)+
M_{A1}^2r\left(v_{2\perp}/v_{1||}\right)^2\sin ^2\theta _{2w} -\left(M_{A1}^2/r\right)\cos ^2\theta _{2w}}\cr
&+&\frac{1}{2}\frac{\left(-M_{A1}\cos\theta _1+\cos 2\theta _1\right)\left\{\left(B_2/B_1\right)\cos\left(\theta _2-\theta _{2w}\right) -
M_{A1}r^{1/2}\left(v_{2\perp}/v_{1||}\right)\sin\theta _{2w} -M_{A1}r^{-1/2}\cos\theta _{2w}\right\} }
{\left(B_2/B_1\right)\cos\left(\theta _2+\theta _{2w}\right)\cos\left(\theta _2-\theta _{2w}\right)+
M_{A1}^2r\left(v_{2\perp}/v_{1||}\right)^2\sin ^2\theta _{2w} -\left(M_{A1}^2/r\right)\cos ^2\theta _{2w}}.
\end{eqnarray}

\section{Cosmic Ray Modified Shock Structure}
We review the structure of cosmic ray modified shocks, largely following \citet{drury81} and \citet{drury83} where fuller details may be found, including
proofs of some of the statements below. These authors concentrated on the case with $\gamma _G=5/3$, $\gamma _{CR}=4/3$, whereas here we
illustrate $\gamma _G=\gamma _{CR}=5/3$, more relevant to the solar case. These hydrodynamic treatments are mainly relevant to quasi-parallel
MHD shocks, where we find the largest effects due to the pre-existing waves. We consider such shocks without these waves, and ignore the effect of the
particle acceleration on the thermal gas, following the references above. The flow obeys the conservation laws for mass, momentum, energy and entropy:
\begin{eqnarray}
\rho v &=& A\cr
Av +P_G + P_{CR} &=& B\cr
\frac{1}{2}v^2 +\frac{\gamma _G}{\gamma _G-1}vP_G +\frac{\gamma _{CR}}{\gamma _{CR}-1}v_{CR}P_{CR}&=&
C +\frac{\kappa}{\gamma _{CR}-1}\frac{\partial P_{CR}}{\partial x}\cr
P_Gv^{\gamma _G} &=& D
\end{eqnarray}
where we have left open the possibility that the cosmic rays flow at a different speed, $v_{CR}$, to the gas, flowing at $v$. Other symbols have their
usual meanings, with the cosmic ray diffusion coefficient given by $\kappa$. We plot the gas propagation through the shock on a pressure-velocity diagram
in Fig. 4. The left panel shows a diagram corresponding to $v_{CR}=v$ above for $\gamma _{CR}=\gamma _G=5/3$. Gas pressure, $P_G$ is on the $y$-axis,
gas velocity $v$ is on the $x$-axis, and the third side of the triangle gives $P_{CR}=B-Av-P_G =0$. The Hugoniot corresponds to the energy
equation with $\partial P_{CR}/\partial x =0$, and for $\gamma _{CR}=\gamma _G$ is the two vertical lines at $v_1$ and $v_2$ as shown, for upstream and downstream respectively. It represents the energetically accessible states of the gas. 

\begin{figure}[t]
\centerline{\includegraphics[width=3.in]{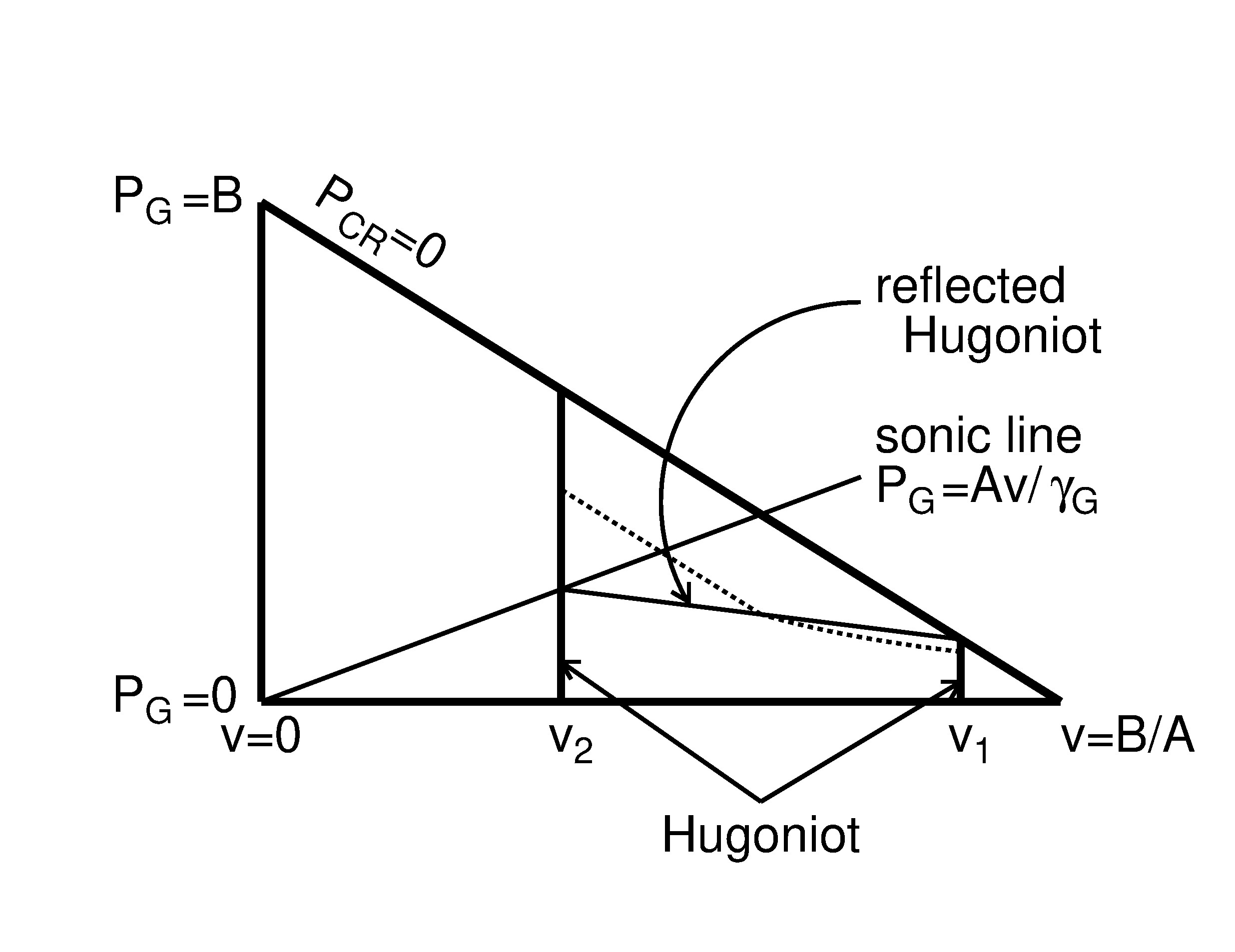}\includegraphics[width=3.in]{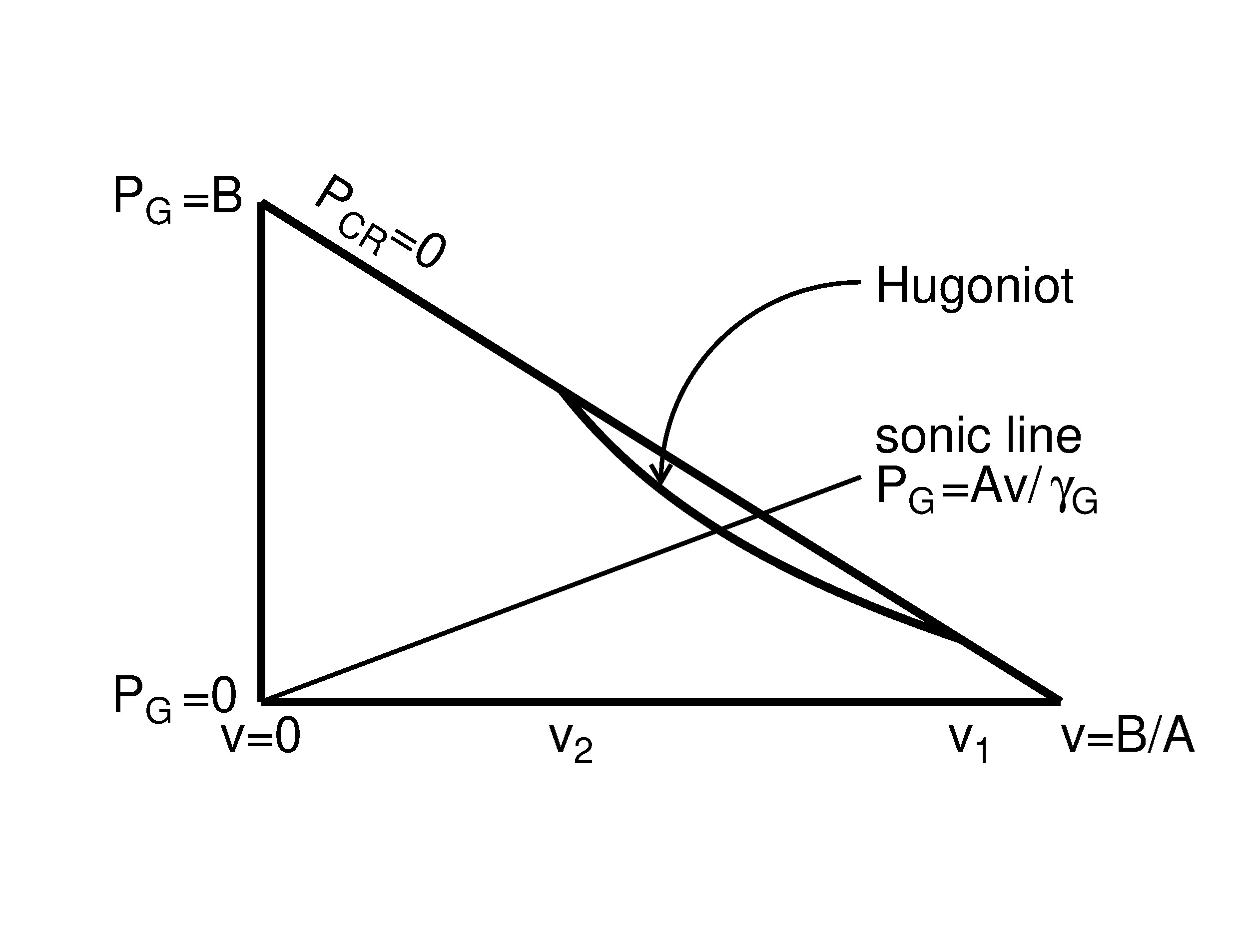}}
\caption{Left: Schematic diagram of shock propagation in the pressure-velocity space for $\gamma _G=\gamma _{CR}=5/3$, $M=1.972$. 
Gas pressure $P_G$ on the $y$-axis, velocity on the 
$x$-axis. Cosmic ray pressure $P_{CR}=0$ delineates the third side of the triangle as shown. The Hugoniot is the two thick vertical lines
at $u_1$ (upstream) and $u_2$ (downstream), and the reflected Hugoniot runs from the intersection of the downstream Hugoniot
and the sonic line and the intersection of the upstream Hugoniot and the $P_{CR}=0$ line. The dotted curve shows the trajectory of plasma with a nonzero
$P_{CR}$, following the adiabatic gas law from its initial position until it hits the reflected Hugoniot, and then making the shock jump to the downstream
Hugoniot. Right: Schematic diagram of shock propagation in the pressure-velocity space with cosmic rays trapped at the shock, i.e. with zero cosmic ray energy flux. The Hugoniot is now the curve shown, and coincides with the reflected Hugoniot.\label{hugoniots}}
\end{figure}

Where $P_G=0$, $P_{CR}=B-Av$ and after substituting this into the energy equation above we find
\begin{equation} v_{1,2}=\frac{\gamma _{CR}}{\gamma _{CR}+1}\frac{B}{A}\left[1\pm\sqrt{1-\frac{2AC}{B^2}\frac{\gamma _{CR}^2-1}{\gamma _{CR}^2}}\right].
\end{equation}
Similarly, where $P_{CR}=0$, the same equation holds with $\gamma _{CR}\rightarrow\gamma _G$, hence the vertical lines for the Hugoniot. We also have
\begin{eqnarray}
B&=&Av+P_G=Av\left(1+1/\gamma _GM_s^2\right)\cr
C&=&\frac{1}{2}Av^2+\frac{\gamma _GP_Gv}{\gamma _G-1}=\frac{1}{2}Av^2+\frac{Av^2/M_s^2}{\gamma _G-1}=Av^2\left(\frac{1}{2}+\frac{1}{M_s^2\left(\gamma _G-1\right)}\right)
\end{eqnarray}
where $M_s^2=Av/\left(\gamma _GP_G+\gamma _{CR}P_{CR}\right)$.

Taking jump conditions $\left[Av+P_G\right]=\left[P_{CR}\right]=0$ and $\left[Av^2/2 +\gamma _GP_Gv/\left(\gamma _G-1\right)\right]=0$
\citet{drury83} proves that gas discontinuities must be symmetric about the sonic line $P_G=Av/\gamma _G$ in the direction of $P_{CR}=0$, so the reflected Hugoniot show the locus of start points for such discontinuities that end on the downstream Hugoniot.  This conclusions would be slightly changed by the
inclusion of terms in $F_{1,2}$ and $E_{1,2}$ from equations 2 and 4, but only mildly.
Gas at a starting point $\left(v_1,P_{G1}\right)$ evolves 
adiabatically with $P_G\propto v^{-\gamma _G}$ until it hits the reflected Hugoniot, whereupon it jumps along a line parallel to $P_{CR}=0$ to the 
downstream Hugoniot which is the final state of the gas. This is illustrated by the dotted curve on the left panel of Figure 4.
To preserve a discontinuity, we require that
\begin{equation} P_{G2}=P_{G1}\left(v_1/v_2\right)^{\gamma _G} > Av_2/\gamma _G,
\end{equation}
so that the pressure at $v_2$ must be higher than that given by the sonic line. Substituting for $v_1$ and $v_2$ from equations B15, and using 
$AC/B^2\simeq 1/2 +\left(M_s^2\gamma _G\left(\gamma _G-1\right)\right)^{-1}$ for $M_s^2 >>1$ from equation B16 we find
\begin{equation}
\left\{\frac{1+\frac{1}{\gamma _G}\sqrt{1-\frac{\gamma _G+1}{\gamma _G}\frac{2}{M_s^2}}}
{1-\frac{1}{\gamma _G}\sqrt{1-\frac{\gamma _G+1}{\gamma _G}\frac{2}{M_s^2}}}\right\}^{1+\gamma _G} > \frac{M_s^2}{1-N}
\end{equation}
where $N=P_{CR}/\left(P_G+P_{CR}\right)$. Even at $N=0$, equation B18 suggests no discontinuity for $\gamma _G=5/3$ for $M_s > 5.5$. This agrees
with the more general result in \citet{becker01}. We argue that this must mean that the model is being stretched beyond its regime of validity, and question whether
cosmic rays really do stream at the velocity of the background gas as $P_{CR}\rightarrow 0$.

The right hand panel of Figure 4 gives the pressure-velocity diagram for the same shock, but with $v_{CR}=0$, i.e. zero cosmic ray energy flux, appropriate for 
when the energetic particles are still trapped in the shock acceleration process. The Hugoniot now resembles that for standard fluid shocks
\citep[e.g.][]{landau87}, and coincides with the reflected Hugoniot.
Gas states initially on the Hugoniot below the sonic line must make a discontinuous jump parallel to $P_{CR}=0$
to a position on the Hugoniot above the sonic line. There is however a maximum cosmic ray pressure, $P_{CR}$, for which this model can be valid. With
\begin{equation} P_{CR}=B-Av-\frac{\gamma _G-1}{\gamma _G}\left(\frac{C}{v}-\frac{Av}{2}\right)
\end{equation}
the maximum is found at $v=\sqrt{2C/A\times\left(\gamma _G-1\right)/\left(\gamma _G+1\right)}$. i.e. where the Hugoniot intersects the sonic line, with value
\begin{equation}
\frac{P_{CR,max}}{Av} = 1+\frac{1}{\gamma _GM_s^2}-\frac{1}{\gamma _G}\sqrt{\gamma _G^2-1+\frac{2\gamma _G+2}{M_s^2}}.
\end{equation}
As $M_s^2\rightarrow\infty$, $P_{CR,max}/Av =P_{CR,max}/\rho v^2\rightarrow 1-\sqrt{1-1/\gamma _G^2} = 0.2$ for $\gamma _G=5/3$.
At $M_s = 2 - 3$, $P_{CR,max}/\rho v^2 = 0.09 - 0.14$, giving $E_{CR,max}/\rho v^2=1.5P_{CR,max}/\rho v^2$ similar to or greater than values used 
in Figures 2 and 3.

\end{document}